\documentclass[aps, prd, twocolumn, superscriptaddress, nofootinbib, floatfix]{revtex4-2}

\usepackage{graphicx}
\usepackage[colorlinks=true,citecolor=blue,linkcolor=blue]{hyperref}
\usepackage{bm}
\usepackage{upgreek}
\usepackage[dvipsnames]{xcolor}
\usepackage{amsmath}
\usepackage{orcidlink}

\makeatletter
\def\farcs{\hbox{$.\!\!^{\prime\prime}$}}

\makeatother

\setcitestyle{authoryear,round}

\newcommand{\Ho}{H$_0$}

\newcommand{\lensrxj}{RXJ1131$-$1231} 
\newcommand{\lenspg}{PG1115$+$080}
\newcommand{\lenshe}{HE0435$-$1223}
\newcommand{\lenswfi}{WFI2033$-$4723}  
\newcommand{\lenses}{\lenshe, \lenspg, and \lenswfi}
\newcommand{\jwst}{\textit{JWST}}

\newcommand{\cat}{CaT}%
\newcommand{\mnras}{MNRAS}
\newcommand{\aap}{A\&A}
\newcommand{\apjl}{ApJL}
\newcommand{\aj}{AJ}
\newcommand{\pasp}{PASP}
\newcommand{\araa}{ARA\&A}

\newcommand{\kmps}{km s$^{-1}$}

\newcommand{\sigaphe}{$220.1\pm4.2$}
\newcommand{\sigappg}{$244.0\pm5.7$}
\newcommand{\sigapwfi}{$204.6\pm6.1$}

\newcommand{\sigtdche}{$227.0\pm6.3$}    
\newcommand{\sigtdcpg}{$236.0\pm7.1$}    
\newcommand{\sigtdcwfi}{$211.0\pm11.2$}  

\newcommand{\dsighe}{$-6.9\ (0.9\sigma)$}
\newcommand{\dsigpg}{$+8.0\ (0.9\sigma)$}
\newcommand{\dsigwfi}{$-6.4\ (0.5\sigma)$}
\newcommand{\dsigmax}{$0.9\sigma$}          
\newcommand{\avgunctdc}{$3.7\%$}            
\newcommand{\avguncnew}{$2.4\%$}            

\newcommand{\constrdiaghe}{$6.2\%$}   \newcommand{\statbudhe}{$5.7\%$}  \newcommand{\sysbudhe}{$2.4\%$}
\newcommand{\constrdiagpg}{$10.7\%$}  \newcommand{\statbudpg}{$9.9\%$}  \newcommand{\sysbudpg}{$3.8\%$}
\newcommand{\constrdiagwfi}{$8.7\%$}  \newcommand{\statbudwfi}{$7.2\%$} \newcommand{\sysbudwfi}{$4.5\%$}
\newcommand{\constrdiag}{$6$--$11\%$}                 
\newcommand{\statbudrange}{$5.7$--$9.9\%$}
\newcommand{\sysbudrange}{$2.4$--$4.5\%$}

\newcommand{\constroffdiaghe}{$1.19\%$}
\newcommand{\constroffdiagpg}{$1.22\%$}   
\newcommand{\constroffdiagwfi}{$1.15\%$}
\newcommand{\constroffdiag}{$\sim1.2\%$}                

\newcommand{\polyhosthe}{5, 0}    \newcommand{\polyhostpg}{10, 0}   \newcommand{\polyhostwfi}{15, 1}   
\newcommand{\polylenshe}{2, 0}    \newcommand{\polylenspg}{4, 0}    \newcommand{\polylenswfi}{3, 0}   
\newcommand{\threshhe}{2.50}      \newcommand{\threshpg}{2.46}      \newcommand{\threshwfi}{2.54}      
\newcommand{\nbinshe}{20}         \newcommand{\nbinspg}{33}         \newcommand{\nbinswfi}{12}         
\newcommand{\nannhe}{10}          \newcommand{\nannpg}{13}          \newcommand{\nannwfi}{9}           
\newcommand{\ntlensiuhe}{9}       \newcommand{\ntlensiupg}{8}       \newcommand{\ntlensiuwfi}{7}       
\newcommand{\ntlensxslhe}{8}      \newcommand{\ntlensxslpg}{9}      \newcommand{\ntlensxslwfi}{9}      
\newcommand{\nthostiuhe}{10}      \newcommand{\nthostiupg}{10}      \newcommand{\nthostiuwfi}{7}       

\newcommand{\nsyscombos}{48}
\newcommand{\polyhe}{(\polylenshe)}   
\newcommand{\polypg}{(\polylenspg)}
\newcommand{\polywfi}{(\polylenswfi)}
\newcommand{\combyhe}{(2,0), (3,0), (2,1), (4,0)}
\newcommand{\combypg}{(4,0), (5,0), (3,0), (4,1)}
\newcommand{\combywfi}{(3,0), (4,0), (2,0), (3,1)}
\newcommand{\gridlensadd}{1--7}   \newcommand{\gridlensmult}{0--3}
\newcommand{\gridhosthe}{3--7$\,\times\,$0--2}
\newcommand{\gridhostpg}{8--12$\,\times\,$0--3}
\newcommand{\gridhostwfi}{15--17$\,\times\,$1--2}

\newcommand{\hostscorehe}{1.72}    
\newcommand{\hostscorepg}{8.18}    
\newcommand{\hostscorewfi}{19.9}   

\newcommand{\vsysdhe}{$-38.6\pm3.2$}
\newcommand{\vsysdpg}{$-138.3\pm5.7$}
\newcommand{\vsysdwfi}{$+192.6\pm6.4$}
\newcommand{\zfiddhe}{0.4546}  \newcommand{\zfidshe}{1.693}
\newcommand{\zfiddpg}{0.311}   \newcommand{\zfidspg}{1.722}
\newcommand{\zfiddwfi}{0.6575} \newcommand{\zfidswfi}{1.662}
\newcommand{\dzdhe}{-0.0002}   \newcommand{\zdnewhe}{0.4544}
\newcommand{\dzdpg}{-0.0006}   \newcommand{\zdnewpg}{0.3104}
\newcommand{\dzdwfi}{+0.0011}  \newcommand{\zdnewwfi}{0.6586}

\newcommand{\vhosthe}{$+419.5\pm21.4$}   
\newcommand{\vhostpg}{$+1543.8\pm10.3$}  
\newcommand{\vhostwfi}{$+7.8\pm14.0$}    
\newcommand{\vemhe}{$+342.4\pm3.7$}      
\newcommand{\vempg}{$+1411.7\pm7.0$}     
\newcommand{\vemwfi}{$+8.6\pm5.4$}       
\newcommand{\dzshe}{+0.0031}   \newcommand{\zsnewhe}{1.6961}   
\newcommand{\dzspg}{+0.0128}   \newcommand{\zsnewpg}{1.7348}   
\newcommand{\dzswfi}{+0.00008} \newcommand{\zsnewwfi}{1.6621}  

\newcommand{\rembinspg}{29 and 31}
\newcommand{\sigbinpgxxix}{$286\pm24$}   
\newcommand{\sigbinpgxxxi}{$65\pm29$}    

\newcommand{\lamrhe}{$0.102\pm0.016$}   
\newcommand{\lamrpg}{$0.323\pm0.037$}   
\newcommand{\lamrwfi}{$0.087\pm0.014$}  
\newcommand{\epshe}{0.144}  \newcommand{\epspg}{0.060}  \newcommand{\epswfi}{0.258}
\newcommand{\rmaxrehe}{0.28} \newcommand{\rmaxrepg}{1.00} \newcommand{\rmaxrewfi}{0.23}
\newcommand{\pgrotmedhl}{86}    
\newcommand{\pgrotmedother}{45} 

\newcommand{\shajiblsft}{\citet{Shajib25a}}
\newcommand{\shajibraccoon}{\citep{Shajib25b}}

\begin{document}

  \title{TDCOSMO XXIX: JWST/NIRSpec IFU Spatially Resolved Kinematics of Three Time-delay Lenses}

  \author{Shawn~Knabel\,\orcidlink{0000-0001-5110-6241}}
  \email{shawnknabel@astro.ucla.edu}
  \affiliation{Department of Physics and Astronomy, University of California, Los Angeles, CA 90095, USA}

  \author{Pritom~Mozumdar\,\orcidlink{0000-0002-8593-7243}}
  \affiliation{Department of Physics and Astronomy, University of California, Los Angeles, CA 90095, USA}

  \author{Anowar~J.~Shajib\,\orcidlink{0000-0002-5558-888X}}
  \affiliation{Department of Astronomy \& Astrophysics, University of Chicago, Chicago, IL 60637, USA}
  \affiliation{Kavli Institute for Cosmological Physics, University of Chicago, Chicago, IL 60637, USA}
  \affiliation{Center for Astronomy, Space Science and Astrophysics, Independent University, Bangladesh, Dhaka 1229, Bangladesh}

  \author{Tommaso~Treu\,\orcidlink{0000-0002-8460-0390}}
  \affiliation{Department of Physics and Astronomy, University of California, Los Angeles, CA 90095, USA}

  \author{Devon~M.~Williams\,\orcidlink{0000-0002-8386-0051}}
  \affiliation{Department of Physics and Astronomy, University of California, Los Angeles, CA 90095, USA}

  \author{Michele~Cappellari\,\orcidlink{0000-0002-1283-8420}}
  \affiliation{Sub-Department of Astrophysics, Department of Physics, University of Oxford, Denys Wilkinson Building, Keble Road, Oxford OX1 3RH, UK}

  \author{David~Law\,\orcidlink{0000-0002-9402-186X}}
  \affiliation{Space Telescope Science Institute, 3700 San Martin Drive, Baltimore, MD 21218, USA}

  \author{Simon~Birrer\,\orcidlink{0000-0003-3195-5507}}
  \affiliation{Department of Physics and Astronomy, Stony Brook University, Stony Brook, NY 11794, USA}

  \author{Takahiro~Morishita\,\orcidlink{0000-0002-8512-1404}}
  \affiliation{Astronomical Institute, Graduate School of Science, Tohoku University, 6–3 Aoba, Sendai 980-8578, Japan}

  \author{William~Sheu\,\orcidlink{0000-0003-1889-0227}}
  \affiliation{Department of Physics and Astronomy, University of California, Los Angeles, CA 90095, USA}

  \author{Massimo~Stiavelli\,\orcidlink{0000-0001-9935-6047}}
  \affiliation{Space Telescope Science Institute, 3700 San Martin Drive, Baltimore, MD 21218, USA}

  \date{\today}

  \begin{abstract}
  Spatially resolved stellar kinematics are critical to the high precision achieved by cosmological probes utilizing time delays of strongly lensed quasars. Combined with high-resolution imaging and lens modeling, dynamical models of the 2D resolved kinematics of the deflector galaxy tightly constrain the mass profile and break the mass-sheet degeneracy, in turn providing tight constraints on the Hubble constant when the time delays are included. We extract stellar kinematics of the deflector galaxies in the quadruply lensed quasar systems \lenses\ from \textit{James Webb Space Telescope} Near-Infrared Spectrograph (\jwst-NIRSpec) integral field spectroscopy. The kinematic maps reach average per-bin total (statistical and systematic) uncertainties of \constrdiag, with average bin-to-bin correlated errors of only \constroffdiag. The aperture-integrated velocity dispersions are statistically consistent with the values used in the previous TDCOSMO analysis (all within \dsigmax), with their average uncertainty reduced from \avgunctdc\ to \avguncnew\ owing to improvements in the data reduction and kinematic-extraction methodology. We classify \lenspg\ as a fast rotator, \lenshe\ and \lenswfi\ as slow rotators within the probed radii, and we refine the source redshifts from the kinematics of the lensed host galaxies. Kinematic maps will be combined with time delays, lens models, and line-of-sight convergence estimates to measure cosmological parameters in the upcoming TDCOSMO 2026 milestone publication.
  \end{abstract}
  
  \keywords{gravitational lensing: strong -- Galaxy: kinematics and dynamics -- Galaxies: elliptical and lenticular, cD -- Galaxies: individual: \lenses -- distance scale}

  \maketitle
  
%
\section{Introduction}

The so-called ``Hubble tension'' describes one of the most contested questions of contemporary observational cosmology. It is the statistical difference of $\sim5\sigma$ in the value of the Hubble constant, \Ho, measured by early- and late-Universe probes \citep{Planck-Collaboration18b_results, Riess22, abdalla_22_hubble_tension, divalentino25_hubble_tension}. If this discrepancy is real, the resolution will require new physical models beyond flat $\Lambda$ cold dark matter ($\Lambda$CDM) cosmology. $\Lambda$CDM is built upon a breadth of observational evidence that must be equally well-described by any models that attempt to reconcile measurements of the Hubble constant. Therefore, the tension must be confirmed with statistical rigor by independent observational probes to justify the challenges facing theoretical efforts.

Time-delay cosmography \citep{treu_marshall16_tdcosmography, Treu22, Wong20, Treu23, Birrer24} measures \Ho\ from the delayed arrival time of flux variations passing through different image paths of a multiply imaged source, typically a quasar or supernova \citep{Refsdal64,Kelly23}. The TDCOSMO collaboration has implemented a hierarchical approach to a sample of galaxy-scale lenses to measure \Ho\ to $5\%$ precision \citep{tdcosmo25_milestone}, henceforth TDC-25, leveraging spatially resolved stellar kinematics of the deflector galaxies to constrain the mass-sheet degeneracy \citep[MSD;][]{falco85, schneider_sluse13}, which is the primary limiting factor of precision for time-delay cosmography with lensed quasars \citep{birrer20_tdcosmo_iv,birrer_treu21}.

\citet{shajib23} measured the first spatially resolved stellar kinematics of a time-delay lens using Keck Cosmic Web Imager (KCWI) integral-field spectroscopy (IFS) on the Keck II telescope at W. M. Keck Observatory. The same lens, \lensrxj, was also observed with James Webb Space Telescope Near-Infrared Spectrograph (JWST-NIRSpec;  \citealt{Boker2022}) IFS, reaching unprecedented spatial resolution for stellar kinematics of the central radii of the deflector galaxy \citep{shajib25b_rxj1131_nirspec}, henceforth TDC-XXIV. The 1D radially averaged profiles of both datasets were used jointly in the TDC-25 analysis and drove the improvements to the precision on \Ho. Follow-up JWST-NIRSpec IFS observations of five other time-delay lenses were also presented for the first time in TDC-25 but were limited to aperture-integrated velocity dispersions due to the stage of reduction at the time the TDC-25 analysis was conducted. This paper builds upon the methods of TDC-XXIV to extract 2D spatially resolved kinematics for three of those five time-delay lenses (\lenses) for the 2026 TDCOSMO full analysis, taking advantage of further improvements in the calibration and data reduction techniques of NIRSpec IFS data.

The paper is organized as follows.
In Section~\ref{sec:observation_data}, we describe the observational program, data acquisition, and reduction.
In Section~\ref{sec:ancillary}, we describe the additional data and information required to extract kinematics. In Section~\ref{sec:methods}, we outline the methods for kinematic extraction. In Section~\ref{sec:result}, we present results. Finally, in Section~\ref{sec:conclusion}, we summarize and conclude the paper.

\section{Observations and data reduction}
\label{sec:observation_data}

In this section, we first provide a brief description of the lenses \lenses\ in our sample in Section~\ref{sec:lens_description}. Then we describe the spectroscopic observation with \jwst-NIRSpec in Section~\ref{sec:jwst_spectra} and the data reduction procedure in Section~\ref{sec:reduction}.

\subsection{Description of the lens systems} \label{sec:lens_description}

The lensed quasar systems \lenses\ are some of the most well-studied objects of their kind. Each has been utilized for time-delay cosmography analyses \citep{treu_koopmans02,Rusu17,Chen19,Bonvin17,birrer20_tdcosmo_iv,rusu20}. See Table~\ref{tab:lens_systems} for references and redshifts.

\begin{table*}[t!]
    \centering
    \begin{tabular}{l | c c c}
        \hline
        & \lenshe\ & \lenspg\ & \lenswfi\ \\
        \hline\hline
        discovery & \citet{wisotzki_02_he0435} & \citet{Weymann80} & \citet{morgan04_wfi2033} \\
        $z_\mathrm{s}$ & \zfidshe\ \citep{Sluse12} & \zfidspg\ \citep{Weymann80} & \zfidswfi\ \citep{Sluse12} \\
        $z_\mathrm{d}$ & \zfiddhe\ \citep{morgan05_he0435_zlens} & \zfiddpg$^{*}$ \citep{kundic97_pg1115,tonry98_pg1115} & \zfiddwfi\ \citep{sluse19_h0licow_x} \\
        time delays & \citet{Bonvin17} & \citet{Schechter97}; \citet{Bonvin18} & \citet{bonvin19_wfi2033} \\
        lens models & Williams et al. (\textit{in prep}) & Williams et al. (\textit{in prep}) & \citet{williams25_wfi2033} \\
        \hline
    \end{tabular}
    \caption{Discovery, redshifts, time-delay measurements, and lens models
    of the three lens systems. The \lenspg\ deflector redshift was found
    independently by the two references given; $^{*}$the value \zfiddpg\ is
    the group center. Lens models are based on JWST-NIRCam imaging.}
    \label{tab:lens_systems}
\end{table*}

\subsection{\jwst-NIRSpec spectroscopy} \label{sec:jwst_spectra}

The \jwst-NIRSpec IFS for \lenses\ was obtained through Cycle 1 program
JWST-GTO-1198 (PI: Stiavelli):
\begin{itemize}
    \item \textbf{Observation dates:} September 9, 2022 (5.06 hours program
    time including overheads; \lenswfi), January 1, 2023 (5.53 hours;
    \lenshe), and April 26, 2023 (4.99 hours; \lenspg).
    \item \textbf{Instrument configuration:} G140M grating with F100LP
    filter, covering an observed wavelength range of 0.97--1.84 $\mu$m with a
    nominal resolution $R \sim 1000$. At the redshifts of the foreground and
    background galaxies, prominent stellar absorption features from rest frame
    near-infrared (NIR, primarily calcium II triplet, henceforth denoted ``\cat'', in the case of the
    foreground lensing deflector) and optical (primarily Ca II H and K, henceforth denoted ``CaHK'', in the
    case of the background source quasar host galaxy) wavelengths were
    observed in the wavelength range.
    \item \textbf{Dithering:} four-point pattern, covering the entire lens
    systems and lensed quasar images; individual observations have slightly
    different fields of view.
    \item \textbf{Readout and exposures:} ``IRS2'' readout mode with
    ``NRSIRS2'' readout pattern. For each dither position, we took two
    integrations of 1750.7~s each, with 24 groups per integration for a total
    of $\sim3.9$~hours for each object.
    \item \textbf{Not obtained:} Micro-Shutter Assembly (MSA) leakage calibration exposures or
    background exposures, to save overhead time.
\end{itemize}


\subsection{Data reduction}
\label{sec:reduction}
 
Turning the raw NIRSpec IFU exposures into suitable data cubes for kinematic extraction at the precision required for time-delay cosmography requires exceptional care beyond the standard software. We modify the custom data reduction pipeline \textsc{RegalJumper}\footnote{\url{https://github.com/ajshajib/regaljumper}}, which was presented by TDC-XXIV and is built on the standard JWST data reduction pipeline\footnote{\url{https://jwst-pipeline.readthedocs.io}; we use version 2.0.0. We use \jwst\ Calibration References Data System context 1535.} \citep{bushouse26_jwst_drp_v2p0p0}.
For the sake of reproducibility, our pipeline will be made publicly available on GitHub upon publication.

\textbf{\textit{Standard processing.}} Our reduction follows the standard three-stage procedure of the JWST pipeline; the custom steps listed below are inserted between and after these stages.
\begin{itemize}
    \item \textbf{Stage~1:} Detector-level corrections generate countrate images per uncalibrated exposure (i.e., \texttt{rate} files from \texttt{uncal} files)\footnote{Group scale correction, data quality initialization, saturation detection, superbias subtraction, reference pixel correction, linearity correction, dark current subtraction, jump detection with default settings for CR shower flagging, ramp fitting, and gain scale correction.}.
    \item \textbf{Stage~2:} Additional instrument-level and observing-mode corrections produce flux-calibrated exposure files (i.e., \texttt{cal} files)\footnote{Assigning the world coordinate system (WCS), source type determination, flat-field correction, path-loss correction, and photometric calibration.}.
    \item \textbf{Stage~3:} A 3D data cube is constructed from the calibrated exposures. Prior to cube building, the pipeline runs the adaptive trace model \citep[ATM;][]{law_clarke26_atm}, new to version 2.0.0, to oversample the detector to a finer grid. This method mitigates ``wiggle'' artifacts due to undersampling of the PSF \citep{Law23, Perna23}. The pixel replacement step recovers pixels flagged {\sc DO\_NOT\_USE} in the oversampled detector exposures before cube building. The data-quality (DQ) array is collapsed bitwise-OR when the cube is built.
\end{itemize}

\textbf{\textit{Configuration choices.}} We implement the following choices in the standard pipeline stages:
\begin{itemize}
    \item \textbf{Stage~1:} We process NRS1 exposures only, which contain nearly all the usable dispersed light; the NRS2 detector adds only noise for our observational configurations.
    \item \textbf{Stage~2:} We skip \textsc{clean\_flicker\_noise} and the background steps (\textsc{bkg\_subtract}, \textsc{master\_background}), replaced by the custom $1/f$ cleaning and post-cube background subtraction below; the MSA leakage steps, because the dedicated leakage calibrations and background exposures were not obtained for this program; and the per-exposure \textsc{cube\_build} and \textsc{extract\_1d}, since cube building is done in Stage~3.
    \item \textbf{Stage~3:} ATM is set to $3\times$ oversampling, and we run pixel replacement with the minimum gradient algorithm. Cubes are built with the drizzling method, centered on the deflector galaxy and aligned to the IFU slicer, at two spaxel scales for testing (0\farcs10 and 0\farcs05); the 0\farcs05 reduction is adopted as the fiducial data cube for kinematics extraction.
\end{itemize}
 
\textbf{\textit{Custom steps.}} Our custom pipeline includes additional steps to clean up $1/f$ noise, cosmic rays (CRs), and artifacts with more flexibility and detailed diagnostics at each step. Compared with the procedure presented by TDC-XXIV and in {\sc RegalJumper}, our modified pipeline skips some steps and introduces new ones. In order of execution:
\begin{enumerate}
    \item \textbf{\textit{Outlier flagging after Stage~1.}} We implement a median sigma clip outlier cleaning of each \texttt{rate} file using a kernel of size (1, 11), i.e., per row along the wavelength dispersion direction. Positive and negative outliers at $>2\times$ or $<1/2\times$ the local running median are flagged as bad pixels, as well as their four orthogonal neighbors. A 0.2 DN/s guard prevents a near-zero baseline from incorrectly flagging noise.
    \item \textbf{\textit{Cosmic-ray cleaning with trace protection after Stage 1.}} We perform additional CR cleaning using the Python package\footnote{\url{https://www.astro.yale.edu/dokkum/lacosmic/}} \textsc{lacosmic} \citep{vanDokkum01}, processing only DQ = 0 (Good) or 4 (Jump-alone). We protect bright traces associated with astrophysical sources with a mask so \textsc{lacosmic} does not mistake the sharp cross-dispersion profile for CRs. For each detector column (cross-dispersion direction), we compute the median and median absolute deviation (MAD) of pixels that are in the $\sim30$ slices spanned by each column. Bright pixels are identified at $3\sigma$ above the per-column median, with $\sigma\sim 1.4826\times\mathrm{MAD}$. A bright pixel is considered associated with an astrophysical source if, within a sliding window of 51 pixels in the row, the fraction of bright pixels is greater than or equal to 20\%. We bridge dim gaps up to 25 pixels wide with binary closing to ensure a smooth contiguous trace and extend the mask to protect $\pm2$ pixels in the cross-dispersion direction to cover PSF wings. This keeps isolated hot pixels for CR cleaning while protecting the continuum of a real source. Any CRs landing on bright pixels are caught by Stage~1 ramp jump detection and Stage~3 cross-dither outlier detection.
    \item \textbf{\textit{$1/f$ noise cleaning before Stage 2.}} We clean the $1/f$ noise using the software package\footnote{\url{https://github.com/JWST-Templates/NSClean/}} \textsc{NSClean} \citep{Rauscher24}. We run a thin pre-pass of Stage~2 (\textsc{clean\_flicker\_noise}, \textsc{bkg\_subtract}, \textsc{master\_background}, \textsc{cube\_build}, and \textsc{extract\_1d} are all skipped) and use the preliminary files for their per-exposure NaN-footprint to help define the background-pixel map for NSClean. We also use a hand-drawn trace mask\footnote{included in the GitHub repository of {\sc RegalJumper}} to achieve a robust cleaning, as recommended by \citet{Rauscher24}. We add ``snowball'' regions detected previously in this stage to this mask for enhanced robustness.
    \item \textbf{\textit{Outlier flagging after Stage~2.}} We manually detect and flag outliers and their adjacent pixels again, as after Stage~1, as a final pass to remove lingering hot pixels, outputting cleaned \texttt{cal} files.
    \item \textbf{\textit{Wiggle correction after cube building.}} We correct any residual ``wiggles'' in the spectra with the software package \textsc{raccoon}\footnote{\url{https://github.com/ajshajib/raccoon}} \shajibraccoon. We assess all pixels within a circular region centered close to the deflector that extend beyond the background source arcs in every direction, correcting pixels that need them in a window of 9703--10800\AA. The correction is very small for our fiducial reductions because ATM already accounts for most of the wiggles.\footnote{{\sc raccoon} settings: detection threshold 0.01, variance-ratio threshold 0.005, 7 amplitudes, 10 frequencies, aperture radius 1 pixel, and comparison aperture circular radius of 3 pixels.}
    \item \textbf{\textit{Background subtraction for final cube.}} We use the JWST background tool to generate a background spectrum to subtract from each of the wiggle-corrected \texttt{wiggleclean} data cubes, producing the final \texttt{bkgsub\_wiggle} data cubes for kinematic analysis.
\end{enumerate}

Intermediate data products and per-step diagnostics are produced, including extracted spectra from central spaxels, white-light images and masks, comparisons of cubes across reduction variants (below) and stages. We produce per-exposure diagnostics including a map of jumps, ramp residuals, a breakdown of pixel DQs, and log-scaled rate image.
 
\textbf{\textit{Pixel replacement and variant data cubes.}} Compared with the data for \lensrxj, the data for the program presented in this paper include more spurious defects that are not fully accounted for in the other cleaning steps, so the pixel replacement step deserves more care. The main algorithms for pixel replacement in the standard JWST pipeline are adjacent profile approximation and minimum gradient estimator (default in the pipeline when \lensrxj\ was reduced). These two algorithms work differently and are subject to different weaknesses when dealing with specific types of defects. For example, the minimum gradient estimator is significantly more local in scope than the profile approximation and thus is much more likely to reject adjacent pixels with very steep gradients. The ATM model improves the pixel replacement compared to previous reduction pipelines, and we consider the ATM reductions with minimum gradient estimator to be the best reduction.

To confirm that this choice does not propagate into the velocity dispersion, we rebuilt the cubes as variants with the ATM step disabled with both minimum gradient and profile approximation pixel replacement. The {\sc raccoon} model and background subtraction were applied to each of the different pixel replacement variants separately, and the {\sc raccoon} configuration was adjusted for the non-ATM variants\footnote{Non-ATM settings: detection threshold 0.3, variance-ratio threshold 0.1; other settings identical.}. We also produced a background-subtracted spectrum from each of the raw (pre-{\sc raccoon}) data cubes to confirm that the wiggle correction step did not bias the absorption features and resulting kinematics in the fiducial ATM reduction. We visually checked for residual wiggles in the signal after correcting with the raccoon model. The ATM reduction appears to consistently remove the majority of the wiggle signal, and there is less residual wiggle in the wiggle-corrected reduction compared with the raw cube and with the other variants. 

We refit all variants with identical machinery; all agree to well within the statistical and fit-systematic uncertainties. We thoroughly tested for differences in the resulting kinematics as a result of these choices in reduction and found them to be a sub-percent effect. In the final product, only the ATM reductions are used. Two further conservative tests---which exposed a failure mode of dropped-pixel reductions---are described more fully in Appendix~\ref{app:pixel_replacement}.

\section{Ancillary ingredients for kinematic extraction}
\label{sec:ancillary}

\subsection{Line spread function}
\label{sec:lsf}

We use the wavelength-dependent formula for the instrumental dispersion $\sigma_{\rm inst}$ assuming a Gaussian line spread function (LSF) provided by \shajiblsft\, which was derived by fitting narrow emission lines from the planetary nebula SMP LMC 58 observed with the same grating and filter configuration of NIRSpec (Program \jwst-CAL-1492, PI: T.~Beck). 

For the deflector fits, we calculate the FWHM at the observed wavelength of the \cat\ (corresponding to rest-frame 8500\AA). For the background host galaxy fits, we calculate it at the observed wavelength of the CaHK (corresponding to rest-frame 3950\AA). We show the values in terms of the FWHM and instrumental dispersion in $\rm km \ s^{-1}$ in Table~\ref{tab:lsf}.

\begin{table*}[t!]
    \centering
    \begin{tabular}{l | c c c}
        \hline
        & \lenshe\ & \lenspg\ & \lenswfi\ \\
        \hline\hline
        \multicolumn{4}{l}{\textit{Spectral fitting}} \\
        fiducial redshifts ($z_\mathrm{d}$, $z_\mathrm{s}$) & \zfiddhe, \zfidshe & \zfiddpg, \zfidspg & \zfiddwfi, \zfidswfi \\
        lens fit window (lens rest) $[\mathrm{\AA}]$          & 8400--8750 & 8400--8750 & 8400--8750 \\
        source fit window (source rest) $[\mathrm{\AA}]$      & 3750--5050 & 3850--5150 & 3850--5050 \\
        polynomial degrees, source fit (add., mult.)          & (\polyhosthe) & (\polyhostpg) & (\polyhostwfi) \\
        polynomial degrees, lens fit (add., mult.)          & (\polylenshe) & (\polylenspg) & (\polylenswfi) \\
        outlier threshold                                      & $\threshhe\sigma$ & $\threshpg\sigma$ & $\threshwfi\sigma$ \\
        quasar images used                                     & all & A1, A2, B & A1, B, C \\
        \hline
        \multicolumn{4}{l}{\textit{Spatial binning}} \\
        s$S/N$ band (lens rest) $[\mathrm{\AA}]$              & 8700--8830 & 8700--8800 & 8700--8830 \\
        target s$S/N$ $[\mathrm{\AA^{-1/2}}]$                 & 45 & 35 & 40 \\
        $N_\mathrm{bins}$ ({\sc PowerBin})                    & \nbinshe & \nbinspg & \nbinswfi \\
        $N_\mathrm{bins}$ (annular)                           & \nannhe & \nannpg & \nannwfi \\
        \hline
        \multicolumn{4}{l}{\textit{Instrumental resolution: FWHM$_\mathrm{inst}$ $\left[\mathrm{\AA}\right]$ / $\sigma_\mathrm{inst}$ $[\mathrm{km\,s^{-1}}]$}} \\
        at Ca\,\textsc{ii} triplet ($\lambda_\mathrm{rest}=8500\,\mathrm{\AA}$) & 12.10 / 124.5 & 12.28 / 140.3 & 11.90 / 107.5 \\
        at Ca\,\textsc{ii} H\&K ($\lambda_\mathrm{rest}=3950\,\mathrm{\AA}$)   & 12.37 / 148.0 & 12.35 / 145.9 & 12.40 / 150.1 \\
        \hline
        \multicolumn{4}{l}{\textit{Total templates: lens / source component}} \\
        lens fit window, Indo-US                           & 727 / 978 & 351 / 978 & 727 / 855 \\
        source fit window, Indo-US                        & $-$ / 978 & $-$ / 978 & $-$ / 855 \\
        lens fit window, XSL                               & 485 / 461 & 488 / 461 & 488 / 456 \\
        source fit window, XSL                             & 485 / 461 & $-$ / 461 & 487 / 456 \\
        \hline
        \multicolumn{4}{l}{\textit{Templates with weight $>1\%$ of the total in the global template}} \\
        lens global template, Indo-US                      & \ntlensiuhe  &  \ntlensiupg & \ntlensiuwfi \\
        lens global template, XSL                          & \ntlensxslhe  &  \ntlensxslpg & \ntlensxslwfi \\
        host global template, Indo-US                      & \nthostiuhe & \nthostiupg & \nthostiuwfi \\
        \hline
    \end{tabular}
    \caption{Per-object analysis settings and spectral inputs. 
    Fit windows are selected independently for each object;
    polynomial degrees are selected by the stability
    criterion of Section~\ref{sec:poly_selection}; the
    s$S/N$ band, redward of the Ca\,\textsc{ii} triplet, is used only to define
    the adaptive binning (Section~\ref{sec:binning}); outlier thresholds follow the
    expected-count criterion of Section~\ref{sec:spikes}. 
    Excluded quasar images for \lenspg\ (image C) and \lenswfi\ (image A2) show reduction
    defects. Instrumental resolution is evaluated in the observed frame at the
    listed rest wavelengths. Template library resolutions: Indo-US
    FWHM $=1.35\,\AA$, i.e.\ $\sigma_\mathrm{temp}=20$/$43$ \kmps\ at the
    Ca\,\textsc{ii} triplet/H\&K \citep{Beifiori11}; XSL $0.74$/$0.40\,\AA$,
    i.e.\ $11$/$13$ \kmps\ \citep{verro22_xshooter}. For \lenshe\ and \lenspg, where a library provides lens-component templates
    in both fit windows, the two sets are synchronized to a common template list.}
    \label{tab:lsf}
\end{table*}

\subsection{Deblended lens light} \label{sec:lens_modeling}

Lens modeling for each of the three objects was conducted on JWST NIRCam imaging by \citet[\lenswfi]{williams25_wfi2033} and Williams et al. (\lenshe\ and \lenspg; \textit{in prep}). The deblended light components allow us to properly scale the light profile for our wavelength-integrated ``white light" data cube image, which we use to estimate the S/N and luminosity-weighted characteristics of the foreground lens galaxy.

\section{Extraction of the kinematic maps}
\label{sec:methods}

In this section, we describe our methodology to extract stellar kinematics of the foreground deflector from the \jwst-NIRSpec data cubes, building upon the methods of TDC-XXIV. 
Velocity dispersions for the background source galaxies are measured only for the sake of removing their contaminating flux from the foreground deflector fits and are not reported here. The spatially resolved kinematics from the background source for the explicit purpose of their study is beyond the scope of this paper and will be reported in a forthcoming paper.

We follow the methodological steps outlined by \citet{knabel_mozumdar25_tdcxix} (henceforth TDC-XIX) to estimate systematic uncertainties and covariance between bins to achieve the precision necessary for time-delay cosmography. We utilize the \textsc{squirrel} pipeline developed and presented in TDC-XXIV,\footnote{\url{https://github.com/ajshajib/squirrel}} which is built on the penalized PiXel Fitting (\textsc{pPXF}) software package\footnote{\url{https://pypi.org/project/ppxf}} \citep{cappellari04_ppxf, Cappellari17, cappellari23_ppxf}.

\subsection{Absorption and emission features}
\label{sec:line_features}
The spectra are all converted to the rest-frame of the background source for
consistency, as described by TDC-XXIV.

\textbf{\textit{Deflector fit.}} The spectrum of the foreground deflector is
fit in a narrow rest-frame NIR wavelength range of $8400$--$8750$~\AA:
\begin{itemize}
    \item \textbf{Target features:} the \cat\ ($\lambda\lambda$8498, 8542,
    8662) absorption lines, plus nearby Ti I $\lambda8435$ and several of a
    complex of Fe I lines present in the fit. The \cat\ lines are an accurate
    probe of stellar kinematics in early-type galaxies (ETGs) \citep{Barth02}; the narrow range
    isolates the stellar kinematics carried in the \cat\ from contamination
    from the background source.
    \item \textbf{Excluded features:} Na I $\lambda8190$, Mg I $\lambda8807$,
    and other absorption features in the observed range outside the fitted
    range. We choose to truncate the fit range because the fits suffered from
    other systematics when the wavelength range was extended significantly
    beyond the \cat\ and surrounding continuum.
    \item \textbf{Overlapping emission lines:} He II $\lambda4687$ (\lenshe), H$\delta$ (\lenspg), and [Fe XIV] $\lambda5304$, He II $\lambda5413$ (\lenswfi). The \lenswfi\ window also covers a complex of quasar Fe II lines, which is handled with a fixed quasar template (Section~\ref{sec:quasar_template}). Lines that are not evident in the source host arcs or quasar spectra are not included; the others were tested as narrow components with kinematics fixed from the source host fit. In all cases, the tested emission lines converged to negligible amplitude or had negligible effect on the binned kinematics. In the end, no emission line is included as a free component in the fiducial deflector fits.
\end{itemize}
\textbf{\textit{Background-source fit.}} The background source quasar host
galaxy is fit in a broad rest-frame optical range of $\sim3850$--$5050$~\AA (per-object windows in Table~\ref{tab:lsf}):
\begin{itemize}
    \item \textbf{Stellar absorption:} CaHK ($\lambda\lambda3934,3969$),
    G-band, Balmer lines, Ca I $\lambda4227$, and Fe I $\lambda4384$ (the
    strongest of many Fe I lines in this region).
    \item \textbf{Expected emission:} lines from the background galaxy and
    quasar across the Balmer series (H10--H$\beta$ for \lenshe; H8--H$\beta$
    otherwise), [O III] ($\lambda\lambda$4363, 4959, and 5007), [Ne III]
    ($\lambda\lambda3869,3968$), He I ($\lambda\lambda4026,4472$), He II
    ($\lambda\lambda$4687 and 5413), the [S II] $\lambda\lambda4068,4076$
    doublet, and broad $\rm H\beta$ and $\rm H\gamma$ components.
    \item \textbf{Fitted emission:} Balmer H8--H$\beta$, [Ne III]
    $\lambda\lambda3869,3968$, [O III] $\lambda\lambda4959,5007$, and He II
    $\lambda4687$, plus per-object additions: He I $\lambda4026$ (\lenshe),
    [S II] $\lambda4068$ (\lenswfi, treated as a single line), and two broad
    $\rm H\beta$ and two broad $\rm H\gamma$ components (\lenspg, \lenswfi).
    Not all emission lines are significantly present in all three objects.
    All fitted emission lines are fit simultaneously with the host galaxy stellar
    component, with multiple groups of lines kinematically tied through trial
    and error.
\end{itemize}

\subsection{Spatial geometry and binning} \label{sec:binning}

\begin{figure*}[t!]
    \includegraphics[width=\textwidth]{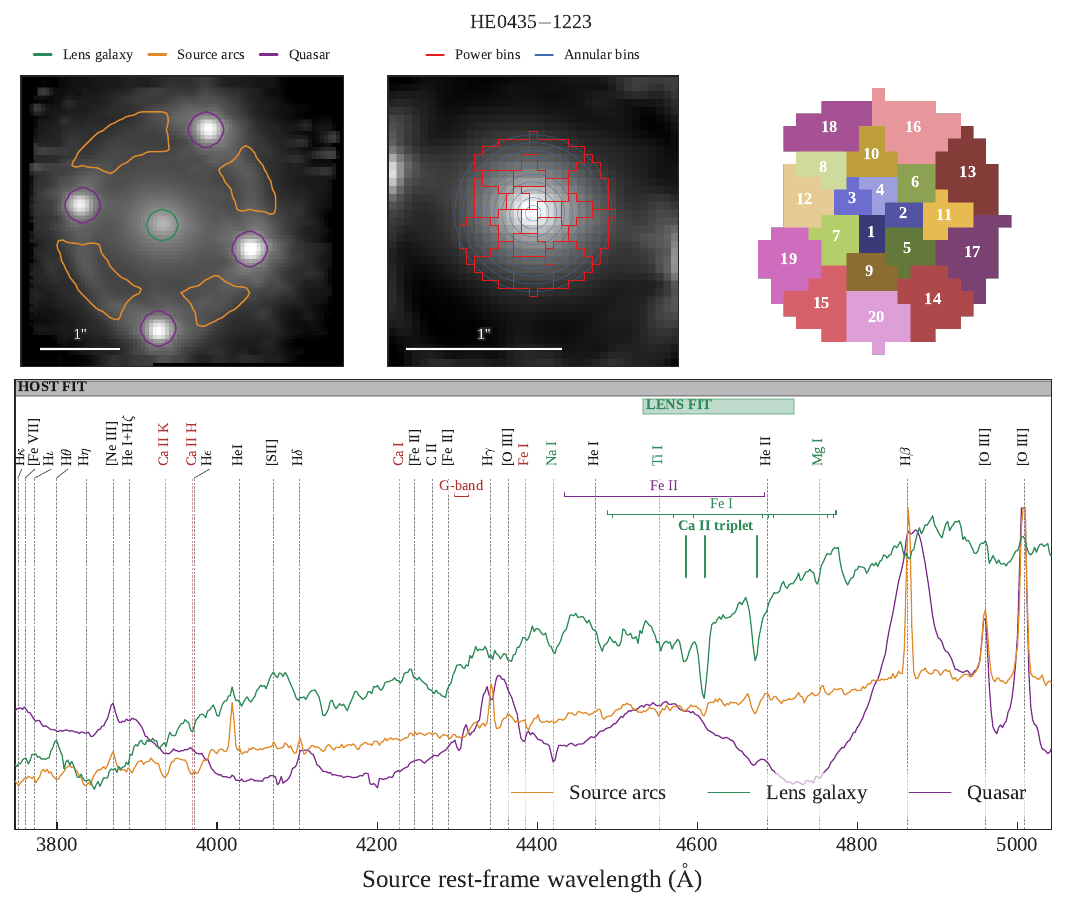}
    \caption{
    Spectral extraction from the data cube for \lenshe\ (published $z_{\rm d} = \zfiddhe$, $z_{\rm s} = \zfidshe$). \textit{Upper left}: White light image integrated over wavelength. Outlined regions show the aperture from which the corresponding spectra in the lower panel are extracted. The white bar indicates the angular scale. \textit{Upper middle}: White light zoomed to $0\farcs9\times0\farcs9$. Outlined regions show the Power bins and annular bins. Kinematics extracted from these bins are shown in Figures~\ref{fig:he0435_maps}--\ref{fig:he0435_cov}. \textit{Upper right}: Power bin map, showing bin numbers that correspond to the spectra shown in Figure~\ref{fig:he0435_bin_fits}. \textit{Lower}: Spectra integrated from apertures shown in upper left panel, in the rest frame of the background source galaxy/quasar. The full wavelength range, highlighted with the gray band, is the range used for the fit to the host galaxy stellar component and emission lines. The green band shows the range fitted for the lens deflector stellar kinematics. Spectral lines are shown with dotted vertical lines, and band features and line complexes are shown with horizontal brackets. Ticks under the horizontal line are centers of lines in the complex. Stellar absorption features for the background host galaxy are marked with red dotted vertical lines and labels. Typical emission lines associated with the host galaxy (and those shared with the quasar spectrum) are labeled in black. Lines and bands associated with the quasar only are shown in purple. Lines and bands associated with the deflector galaxy are shown in green, with the CaT lines displayed with short solid lines for visibility of the line shapes.
    }\label{fig:he0435_comp_fig}
    \end{figure*}

\begin{figure*}[t!]
\includegraphics[width=\textwidth]{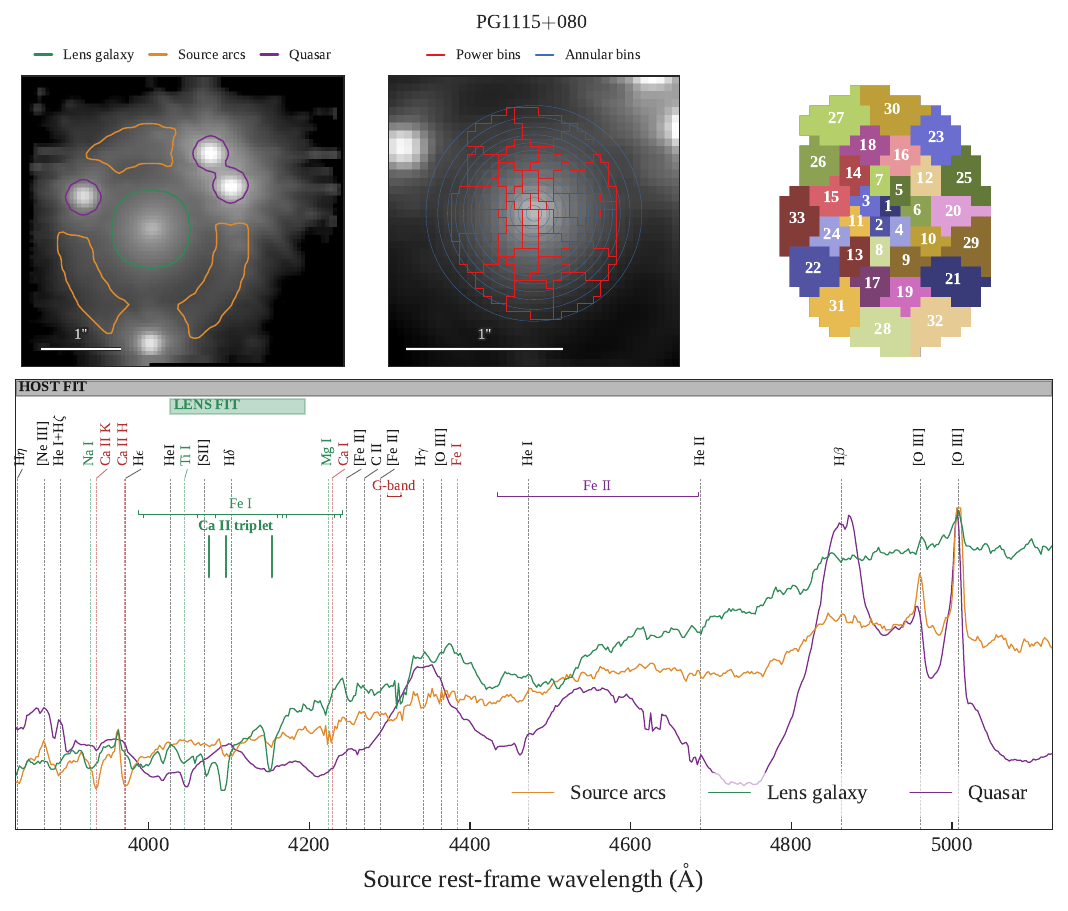}
    \caption{Same as in Figure~\ref{fig:he0435_comp_fig} for \lenspg\ (published $z_{\rm d} = \zfiddpg$, $z_{\rm s} = \zfidspg$). Kinematics extracted from these bins are shown in Figures~\ref{fig:pg1115_maps}--\ref{fig:pg1115_cov}. \textit{Upper right}: Power bin map, showing bin numbers that correspond to the spectra shown in Figures~\ref{fig:pg1115_bin_fits_1}--\ref{fig:pg1115_bin_fits_2}.} \label{fig:pg1115_comp_fig}
\end{figure*}

We first extract spectra from spatial apertures over the data cube to examine the spectral features (as described in Section~\ref{sec:line_features}) of the three components: the central foreground lens galaxy, the lensed arcs of the background source galaxy, and the multiply-imaged quasar. These apertures and extracted spectra for each object are shown in Figures~\ref{fig:he0435_comp_fig}, \ref{fig:pg1115_comp_fig}, and \ref{fig:wfi2033_comp_fig}. These ``pristine'' spectra of each component are used to define the ``global template'' components that are used in the fitting of the spatial bins (see Section~\ref{sec:templates}). Each aperture is individually tuned for each object through careful trial and error with the intent to minimize contamination from other components and maximize the S/N. Blended absorption features from the deflector and source can both be seen in the spectrum of the other. Both deflector and source are fitted during the construction of each global template component, and the raw deflector and source host spectra are never used as fit components. The sensitivity to aperture size is subdominant to statistical uncertainties and systematics arising from the spectral fits.

For appropriate spatial binning of each data cube with the {\sc PowerBin} method\footnote{\url{https://pypi.org/project/powerbin/}} \citep{cappellari25_powerbin}, we first set a target $S/N$ for each bin. We define a ``specific'' $S/N$ (s$S/N$) as $\textrm{s$S/N$} \equiv (S/N) / \sqrt{\Delta L}$, where $S$ is the summed flux within a wavelength range $\Delta L$ and $N$ is the noise summed in quadrature within the same range. The $\sqrt{\Delta L}$ term standardizes the s$S/N$ as it cancels out the improvement in the $S/N$ solely due to an increase in the summed wavelength range. Our target s$S/N$ is individually determined for each object through experimentation to sufficiently cover a useful range of radii from the center of the galaxy and achieve stable systematics while avoiding spurious extreme velocity dispersions that are indicative of reduction defects standing out above the optimal s$S/N$. In practice we start from a target s$S/N\sim40$ and inspect the result. We lower the target if it yields fewer than $\sim$10 bins (we aim for roughly 30), keeping the central bins as small as possible (ideally single spaxels), and raise it if the spectra in the outermost bins are visibly unusable.

For each spaxel within a hand-drawn elliptical mask around the foreground lens galaxy, we estimate the s$S/N$ in a continuum range slightly redward of the \cat. The signal (not the noise) in each spaxel is scaled by the lens light fraction taken from the deblended lens model light from \citet{williams25_wfi2033} and Williams et al. (\textit{in prep}) described in Section~\ref{sec:ancillary} to remove blended background source light that increases the signal of the continuum. Power bins are created from groups of spaxels within the mask that pass a minimum threshold of s$S/N=1$. We also produce concentric, circular annular bins at the spaxel scale for visualization and comparison purposes. These binning schemes are shown in Figures~\ref{fig:he0435_comp_fig}, \ref{fig:pg1115_comp_fig}, and \ref{fig:wfi2033_comp_fig}. Target s$S/N$, wavelength ranges, and numbers of bins for each object are listed in Table~\ref{tab:lsf}.

\subsection{Log-rebinning of spectra} \label{sec:rebinning}

We follow the standard procedure to prepare the spectra for kinematic fitting with \textsc{pPXF} by rebinning the spectra onto a logarithmically sampled common velocity grid \citep{Cappellari17}, such that each rebinned spectrum has the same number of pixels as before rebinning. Each object has a unique velocity scale. {\sc Squirrel} propagates the uncertainties and covariance per-wavelength pixel by creating a covariance matrix using the Monte Carlo method. The spectra are resampled 5000 times from their uncertainties and log-rebinned, from which the covariance is calculated.

\subsection{Polynomial degree selection} \label{sec:poly_selection}

We include additive and multiplicative Legendre polynomials in every spectral fit. The multiplicative polynomial absorbs low-order differences in continuum shape between the data and the templates, while the additive polynomial absorbs residual additive contributions such as imperfect background subtraction and template mismatch \citep{Cappellari17}. The selection of polynomial degrees is a known systematic in kinematic extraction: overly flexible polynomials can absorb template mismatch and bias the measured dispersion. Expanding on the methods described in TDC-XIX, we select the polynomial degrees with an outcome-blind (to the value of $\sigma$ or $\chi^2$) stability criterion, applied identically to the deflector and host fits: we take the lowest-degree cell in a grid of candidate degrees for which the two template libraries agree and the fitted dispersion is insensitive to a one-step change in either degree. The criterion, the grids, and the stability maps for the three objects are given in Appendix~\ref{app:poly_selection}. The selected fiducial degrees are \polyhe, \polypg, and \polywfi\ for \lenshe, \lenspg, and \lenswfi, respectively (Table~\ref{tab:lsf}); the neighboring cells enter the systematics grid (Table~\ref{tab:systematic_choices}).

\subsection{Spectral templates} \label{sec:templates}

We use two empirical stellar template libraries commonly used throughout the literature: the Indo-US \citep{Valdes04} and the X-shooter Spectral Library (XSL) DR3 \citep{verro22_xshooter}. We use only the templates selected and compiled as the ``cleaned" libraries (i.e., suitable for high-precision stellar kinematics) by TDC-XIX and further remove spectra with defects in the fitted wavelength ranges according to the flags described therein.\footnote{The ``cleaned'' libraries and quality flags can be obtained from \url{https://github.com/TDCOSMO/KINEMATICS_METHODS}.} 

We create eight different template sets for each combination of fit window (lens \cat\ and host CaHK), fit component (lens and host), and library (Indo-US and XSL). We show the number of templates that are kept for each combination in Table~\ref{tab:lsf}. 
Lens-component templates are unavailable from the Indo-US library in the host fit window for all three objects, for reasons that differ by object. The host window is defined in the source rest frame, so the lens component templates must cover it at lens rest-frame wavelengths multiplied by $\rm(1+z_s)/(1+z_d)$. For \lenshe\ and \lenswfi\ the mapped windows ($\sim$6900--9350 and $\sim$6200--8100~\AA) lie within the Indo-US wavelength range, but they span wide, telluric-contaminated stretches in which every Indo-US template carries a TDC-XIX quality flag, so all are excluded. For \lenspg\ the ratio is $\sim2.1$, which pushes the requirement to lens rest-frame $\sim$8000--10700~\AA, beyond the red end of the Indo-US library ($\sim$9460~\AA). The XSL lens component is likewise unavailable for \lenspg\ because the mapped window includes the rejected region where the XSL optical and NIR spectral observations were stitched. In the Indo-US host fits, the lens component is therefore taken from the XSL library for \lenshe\ and \lenswfi. For \lenspg, the host fits include no lens stellar component, and the lens continuum, which is primarily smooth throughout the host fit window, is absorbed by the corrective polynomials.

The host fit is primarily used to build the background host stellar component and emission line templates. The resolutions for the JWST-NIRSpec data for each object and for the stellar template libraries in both the lens fit window and source fit window are listed in Table~\ref{tab:lsf}, as well as the total number of useable templates for each library in each fit window for each object. 
As in TDC-XXIV, we consider all templates with weights greater than $1\%$ to contribute meaningfully to the fit, and we report these numbers in Table~\ref{tab:lsf}. 
The stellar template libraries are convolved to the instrumental resolution with a wavelength-dependent Gaussian kernel (the \texttt{varsmooth} routine of \textsc{ppxf} in \textsc{squirrel}), accounting for the difference between the instrumental and template-library line spread functions at each wavelength.

\subsubsection{Template for the host galaxy} 
\label{sec:host_template}

To create a template for the quasar-host galaxy's stellar continuum, we first extract the spectrum from a region containing the lensed arcs, as shown in Figures~\ref{fig:he0435_comp_fig},~\ref{fig:pg1115_comp_fig},~and~\ref{fig:wfi2033_comp_fig}, within a host rest-frame window of $3750$--$5050$~\AA\ (\lenshe), $3850$--$5150$~\AA\ (\lenspg), and $3850$--$5050$~\AA\ (\lenswfi; see Table~\ref{tab:lsf}). Then, the spectra are fit with the Indo-US and XSL template libraries, while also accounting for the contribution from the lens galaxy's stellar continuum and host galaxy emission lines at the corresponding wavelength range, as described in Section~\ref{sec:line_features}. The covariance matrix of the summed arc spectrum is projected onto the nearest positive-definite matrix before fitting.
We select the combination of additive and multiplicative polynomial degrees with the stability criterion described in Section~\ref{sec:poly_selection}, applied to a grid of candidate degrees for each object.
We combine all templates with non-negative weights from the Indo-US library for the best-fit model in a weighted sum to serve as the global template for the host galaxy in later fits.

The velocity of the host stellar component is fixed from this point. Whether the host stellar velocity dispersion is likewise fixed or left free in the deflector fits is decided per object: $\sigma_{\rm host}$ is fixed if and only if it is formally constrained ($\sigma/\delta\sigma \geq 2$) and its stability score at the chosen polynomial combination is $\leq 3$ (a looser tolerance than the pass criterion used for the polynomial selection; Appendix~\ref{app:poly_selection}). For \lenshe\ the score is \hostscorehe\ and $\sigma_{\rm host}$ is fixed; for \lenspg\ (score \hostscorepg) and \lenswfi\ (\hostscorewfi) the fitted $\sigma_{\rm host}$ is polynomial-dependent and is left free in the deflector fits. The effect on the deflector velocity dispersions is negligible in all cases, with differences $\lesssim1$ \kmps\ between fixed and free.

\subsubsection{Templates for the quasar spectra} \label{sec:quasar_template}

\begin{figure*}[t!]
    \includegraphics[width=\textwidth]{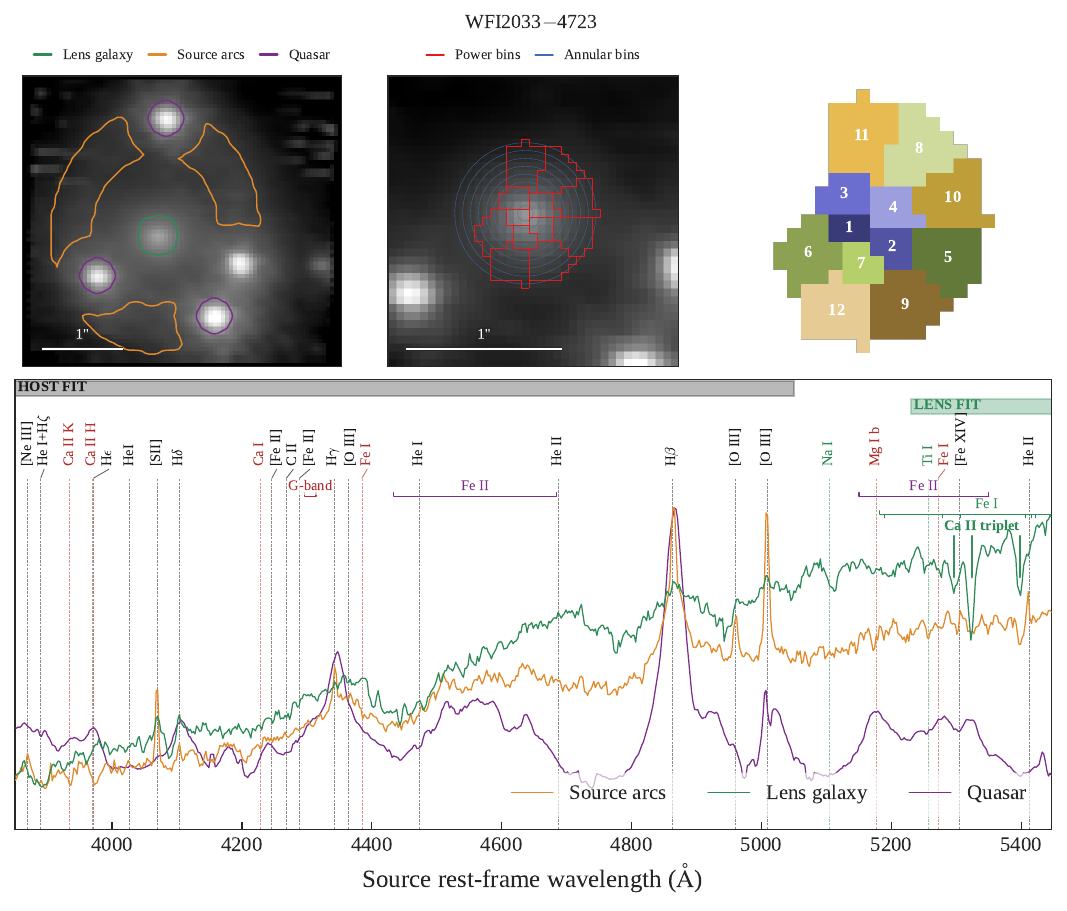}
    \caption{Same as in Figure~\ref{fig:he0435_comp_fig} for \lenswfi\ (published $z_{\rm d} = \zfiddwfi$, $z_{\rm s} = \zfidswfi$). Kinematics extracted from these bins are shown in Figures~\ref{fig:wfi2033_maps}--\ref{fig:wfi2033_cov}. \textit{Upper right}: Power bin map, showing bin numbers that correspond to the spectra shown in Figure~\ref{fig:wfi2033_bin_fits_1}.} \label{fig:wfi2033_comp_fig}
    \end{figure*}

We extract spectra from each of the four background quasar images as shown in the mask in Figures~\ref{fig:he0435_comp_fig}, \ref{fig:pg1115_comp_fig}, and~\ref{fig:wfi2033_comp_fig}. We average over three or four images (purple outlines show the images that are included) to smooth out defects in the continuum. Broad and blended emission lines described in Section~\ref{sec:line_features} are labeled. For \lenshe\ and \lenspg, the quasar spectra are smooth throughout the lens fit wavelength range, so quasar contamination is absorbed by the corrective polynomials. The lens fit window for \lenswfi\ covers a busier region of background source arc and quasar lines, including the complex of Fe II lines that is difficult to model with individual template line components. We use the combined quasar spectrum as shown in Figure~\ref{fig:wfi2033_comp_fig} as a fixed template that is scaled by its weight and not broadened by kinematics.

\subsubsection{Template for the lens galaxy's spectra} \label{sec:lens_template}

We construct the stellar template for the lens galaxy from the spectrum extracted from the deflector mask shown in Figures~\ref{fig:he0435_comp_fig},~\ref{fig:pg1115_comp_fig},~and~\ref{fig:wfi2033_comp_fig} in a lens rest frame wavelength range of $8400$--$8750$~\AA. 
We include the background host stellar template and emission line templates with fixed kinematics, as fitted in Section~\ref{sec:host_template}, together with the quasar components described in Section~\ref{sec:quasar_template}. As shown in that section, the individual emission line components are negligible in the small window probed by the lens fit. As described in Section~\ref{sec:host_template} for the background host galaxy stellar templates, we fit the spectrum over a range of additive and multiplicative polynomials for both the Indo-US and XSL stellar template libraries. For the best polynomial combination, for each template library, we collect all stellar templates with non-negative weights in a weighted sum to be the global template for that library. This global template is used for fitting the kinematics in each spatial bin. For bookkeeping purposes, Table~\ref{tab:lsf} lists the number of templates that receive more than $1\%$ of the total weight in each of these fits.

\subsection{Outlier masking}
\label{sec:spikes}

The spectra contain spurious features that stand out in the residuals and do not correspond to expected absorption or emission lines from the foreground lens or background source galaxies, such as artifacts from uncleaned cosmic rays or poorly recovered pixels from the pixel-replacement algorithms. These can significantly affect the measured kinematics, especially in bins far from the center of the galaxy, and we mask them. Outliers are identified in the noise-normalized residuals of an initial fit with a threshold set so that fewer than one pixel per fit window is expected to exceed it by chance ($\sim2.5\sigma$, Table~\ref{tab:systematic_choices}); the same mask is applied to both template libraries, and the cores of the \cat\ lines are protected from rejection. The procedure, and our reasons for masking these features rather than modeling them as additional emission-line components as in TDC-XXIV, are given in Appendix~\ref{app:spikes}.

\subsection{Estimating systematic uncertainty} \label{sec:systematic}

We follow the procedure developed in TDC-XIX to estimate the systematic effects of several different aspects of the fit, including the stellar template library, the degrees of additive and multiplicative polynomials, the fitted wavelength range, and the threshold of outlier rejection. The \nsyscombos\ combinations are each considered individually. 

For each setup, we first fit the foreground lens aperture spectrum from Figures~\ref{fig:he0435_comp_fig},~\ref{fig:pg1115_comp_fig},~and~\ref{fig:wfi2033_comp_fig} to get a new global template for the setup, which will be applied to all the spatial bins. For this fit, we restrict the libraries to the subsets of templates that were used to construct the fiducial global templates for Indo-US and XSL, but we allow the fit to freely assign new weights to those templates.  We then fit all the spatial bins with the \nsyscombos\ selections (using the global templates for each setup) and marginalize over the results in two stages. For systematics that do not change the number of data points (e.g., template library and polynomial degree) we weight by the Bayesian information criterion (BIC). For the wavelength ranges and outlier rejection thresholds, we combine the resulting kinematic measurements and their covariances with equal weights; we do not use BIC weighting across wavelength ranges, as it would weigh more heavily the case with a smaller number of data points, which is undesirable. 

We list the selections in Table~\ref{tab:systematic_choices} and the resulting bin-averaged error budget in Table~\ref{tab:systematic_error}. We use $\zeta_B = \sqrt{\left\langle C^{\sigma}_{B,ij}/\bar{\sigma}_i\bar{\sigma}_j\right\rangle_{i\neq j}}$, as defined by TDC-XIX, to denote the average off-diagonal correlated errors: $\zeta_B = $ \constroffdiaghe, \constroffdiagpg, and \constroffdiagwfi\ for \lenshe, \lenspg, and \lenswfi, respectively.

\begin{table}[t!]
    \centering
    \begin{tabular}{l|l}
        \hline
        systematic source & choices \\
        \hline
        \hline
        template library            & Indo-US, XSL \\
        polynomial degree           & \lenshe: \combyhe \\
        (add, mult)                 & \lenspg: \combypg \\
                                    & \lenswfi: \combywfi \\
        wavelength range [\AA]      & 8400--8750, 8435--8750, 8400--8715 \\
        rejection threshold         & $2.50/2.46/2.54$,\ $3.0$ \\
        \hline
    \end{tabular}
    \caption{Systematic grid axes marginalized over at the spatial-bin level. Wavelength ranges are in the lens rest frame. The rejection-threshold axis pairs the per-object derived value (listed as \lenshe/\lenspg/\lenswfi; see Table~\ref{tab:lsf}) with a fixed $3.0\sigma$.}
    \label{tab:systematic_choices}
\end{table}

\begin{table}[t!]
    \centering
    \setlength{\tabcolsep}{3pt}
    \begin{tabular}{l|cccc}
        \hline
         & $\langle\delta\bar{\sigma}/\bar{\sigma}\rangle$ & $\langle\Delta_B\bar{\sigma}/\bar{\sigma}\rangle$ & $\langle\delta_{\rm tot}\bar{\sigma}/\bar{\sigma}\rangle$ & $\zeta_B$ \\
        \hline
        \hline
        \lenshe\  & \statbudhe  & \sysbudhe  & \constrdiaghe  & \constroffdiaghe \\
        \lenspg\  & \statbudpg  & \sysbudpg  & \constrdiagpg  & \constroffdiagpg \\
        \lenswfi\ & \statbudwfi & \sysbudwfi & \constrdiagwfi & \constroffdiagwfi \\
        \hline
    \end{tabular}
    \caption{Bin-averaged statistical, systematic, and total marginalized uncertainties on the velocity dispersions, and the average off-diagonal correlated error $\zeta_B$, for each object. Averages are computed over the retained bins (for \lenspg, excluding the two removed bins \rembinspg). The total is the bin average of the per-bin quadrature sum of the statistical and systematic terms.}
    \label{tab:systematic_error}
\end{table}

\subsection{Kinematic fitting procedure summary} \label{sec:procedure_summary}

Here, we summarize the steps involved in our fitting procedure:
\begin{enumerate}
    \item We extract aperture spectra of the deflector, the quasar images, and the background source host lensed arcs, and construct the spatial binning (Section~\ref{sec:binning}).
    \item We log-rebin all spectra and propagate per-pixel covariance via Monte Carlo resampling (Section~\ref{sec:rebinning}).
    \item We fit the arc spectrum to fix the host stellar and emission-line velocities (and, for \lenshe, the host stellar velocity dispersion) and build the global host template (Section~\ref{sec:host_template}), construct the quasar templates (Section~\ref{sec:quasar_template}), and fit the deflector aperture spectrum to build the global lens template for each library (Section~\ref{sec:lens_template}).
    \item We fit each spatial bin with the global templates, identifying and rejecting spurious features by iterative outlier rejection with the \cat-core guard (Section~\ref{sec:spikes}).
    \item We repeat the bin fits over the \nsyscombos\ selections of Table~\ref{tab:systematic_choices}, reweighting the global template for each setup, and combine the results in two steps: first with BIC weights over the template library and polynomial degrees, then with equal weights over the wavelength range and rejection threshold, to obtain the final kinematics and covariances (Section~\ref{sec:systematic}).
\end{enumerate}

\section{Results} 
\label{sec:result}

For each of the three objects we describe the deflector kinematics, including 2D maps, radial profiles, integrated aperture velocity dispersions, covariance, and small corrections to redshifts.  The background host stellar and emission line velocity dispersions will be measured robustly in a forthcoming paper.

Redshift corrections are connected to the mean velocity offsets of the galaxies as $1 + z = (1 + z_{\rm fid})\exp(V/c)$, i.e. $\Delta z = (1 + z_{\rm fid})\left[\exp(V/c) - 1\right]$ \citep[Section~2.3 of][ and Section~2.2 of \citealp{cappellari23_ppxf}]{Cappellari17}. Throughout, $V$ is the mean velocity returned by the aperture-integrated fit relative to the fiducial redshift $z_{\rm fid}$ adopted in Table~\ref{tab:lsf}; we refer to it as the velocity offset rather than the systemic velocity, since it is defined relative to $z_{\rm fid}$. Velocity offsets of the deflector and background-source components, and the
corresponding redshift adjustments, are collected in Table~\ref{tab:velocity_offsets}. For the spatially resolved kinematics, the peculiar velocity of each bin is the difference between its fitted velocity and a robust systemic velocity $V_\mathrm{sys}$ estimated from the velocity map with \textsc{pafit} \citep{krajnovic_06_fitkinpa}, computed as in Section~2.2 of \citet{cappellari23_ppxf}. We consider the emission lines to be the more robust redshift determination than the stellar component. The host stellar offsets are sensitive to the quality of the arc stellar-continuum fit and are not used in any subsequent analysis.

Power bin-averaged statistical, systematic, and total marginalized uncertainties on the velocity dispersions, and the average off-diagonal correlated error $\zeta_B$ are shown in Table~\ref{tab:systematic_error}. We show the white light images with binning and extraction apertures overlaid, maps of the bin numbers, and extracted spectra for each of the lensing deflector, background source host galaxy, and quasar. Each object has distinct, separable components that reveal a wealth of absorption and emission lines. We then show the 2D maps of the velocity dispersion and mean velocity measurements, radial profiles for the 2D maps and annular bin fits, and the covariance matrices for bin fit velocity dispersions.

\subsection{\lenshe}

\begin{figure}[t!]
    \includegraphics[width=\columnwidth]{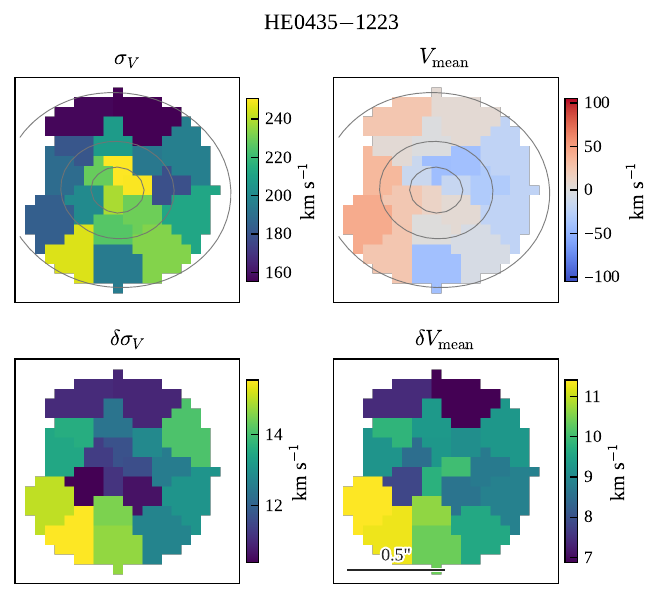}
    \caption{Kinematic maps, showing velocity dispersion and mean velocity measured within each Power bin, with marginalized uncertainties. Gray contours on the $\sigma_V$ and $V_{\rm mean}$ panels show isophotes of the double S\'ersic model of the deblended deflector light, spaced at $0.5$\,mag intervals in surface brightness below the peak. The bar in the $\delta V_{\rm mean}$ panel indicates the angular scale.} \label{fig:he0435_maps}
\end{figure}

\begin{figure}[t!]
    \includegraphics[width=0.9\columnwidth]{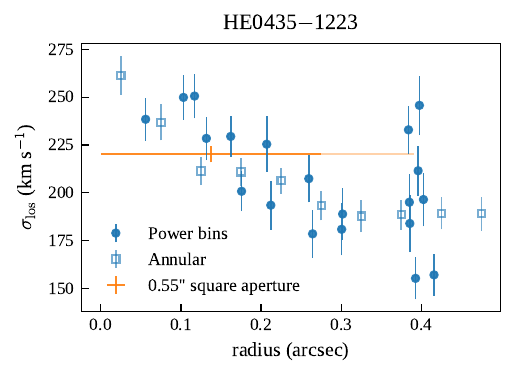}
    \caption{Radial profiles of velocity dispersions for Power bins and annular bins, with marginalized uncertainties.}\label{fig:he0435_radial}
\end{figure}

The foreground deflector galaxy's 2D velocity dispersion map shows a clear peak and symmetrical declining profile, shown in Figure~\ref{fig:he0435_maps}, which is also shown in its annular profile in Figure~\ref{fig:he0435_radial}. The annular profile averages over the bins at similar radius, flattening the profile around 0\farcs4. The per-bin mean velocity is small relative to the velocity dispersion, but there is a clear rotational axis. Mean per-bin statistical errors on velocity dispersions are \statbudhe, and systematic errors are \sysbudhe, for a total added in quadrature of \constrdiaghe. The off-diagonal correlated error is on average \constroffdiaghe, achieving the accuracy required for precision cosmology.

\begin{figure}[t!]
    \includegraphics[width=\columnwidth]
{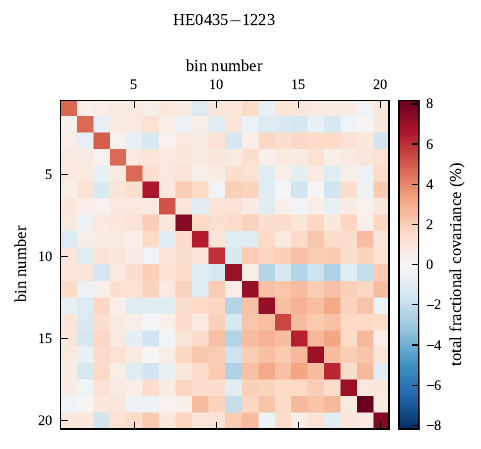}
    \caption{Covariance matrix for Power bin fits shown as a fraction of the bin velocity dispersion, with marginalized uncertainties.} \label{fig:he0435_cov}
\end{figure}

\subsection{\lenspg}

The velocity dispersion profiles in 2D and 1D are symmetric and decline in a clear profile, with the exception of bin 29, which is unreasonably high  (\sigbinpgxxix\ \kmps) with unremarkable statistical uncertainties. The defect persists when combining those spaxels in different ways, and we choose to remove the bin rather than tailor it. Bin 31 falls below the reliable-measurement floor set by the velocity scale (\sigbinpgxxxi\ \kmps) and is likewise removed. Both removed bins are flagged and excluded from the quoted averages. The high bin mean velocities exhibit clear rotation. The lowest-dispersion outer bins lie along the kinematic major axis: their median absolute peculiar velocity $\left\lvert{V - V_\mathrm{sys}}\right\rvert$ is \pgrotmedhl\ \kmps, versus \pgrotmedother\ \kmps\ for the other bins at $r\geq0\farcs27$. Mean per-bin statistical errors on velocity dispersions (excluding bins \rembinspg) are \statbudpg, and systematic errors are \sysbudpg, for a total added in quadrature of \constrdiagpg. The off-diagonal correlated error is on average \constroffdiagpg.

\begin{figure}[t!]
    \includegraphics[width=\columnwidth]{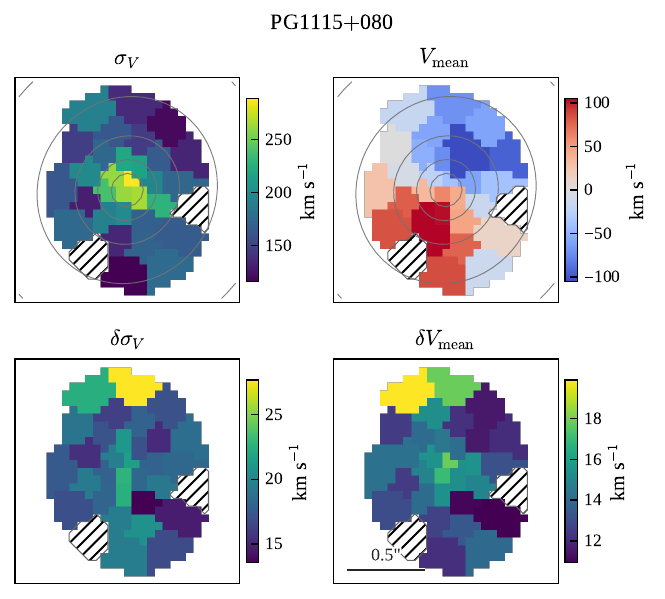}
    \caption{Same as Figure~\ref{fig:he0435_maps}. The two excluded bins (\rembinspg) are marked with hashes.\label{fig:pg1115_maps}} 
\end{figure}

\begin{figure}[t!]
    \includegraphics[width=0.9\columnwidth]{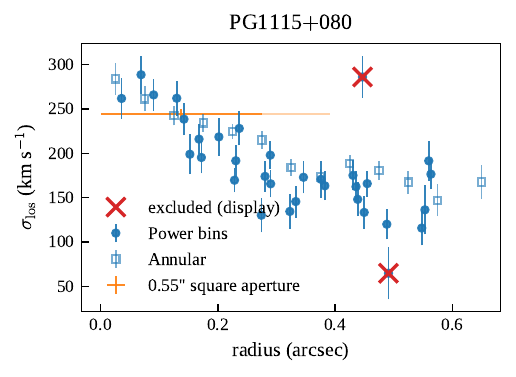}
    \caption{Same as Figure~\ref{fig:he0435_radial}. The two excluded bins (\rembinspg) are marked with red X.\label{fig:pg1115_radial}}      
\end{figure}

\begin{figure}[t!]
    \includegraphics[width=\columnwidth]
{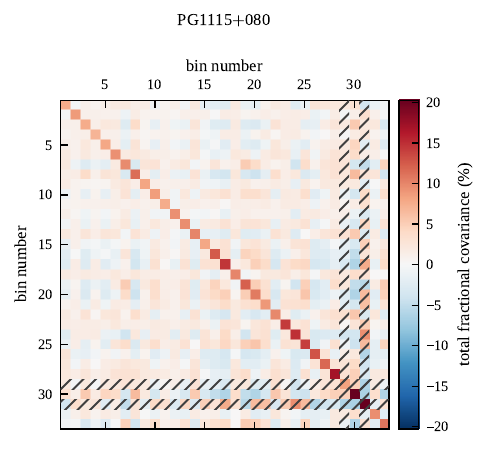}
    \caption{Same as Figure~\ref{fig:he0435_cov}. The rows and columns of the two excluded bins (29 and 31) are hatched and do not set the color scale.\label{fig:pg1115_cov}}
\end{figure}

\subsection{\lenswfi}

The overlap of the quasar Fe II complex with the deflector \cat\ makes the extraction of precise kinematics in the outer bins of this object challenging. In addition, defects that were not easily removed by the pipeline resulted in spaxels that were unusable despite the nominally acceptable s$S/N$, and we remove those before fitting by masking those regions by hand. The velocity dispersion maps and radial profile are still reasonably informative, and the mean velocity map is random with no clear rotational axis. Mean per-bin statistical errors on velocity dispersions are \statbudwfi, and systematic errors are \sysbudwfi, for a total added in quadrature of \constrdiagwfi. The off-diagonal correlated error is on average \constroffdiagwfi.

\begin{figure}[t!]
    \includegraphics[width=\columnwidth]{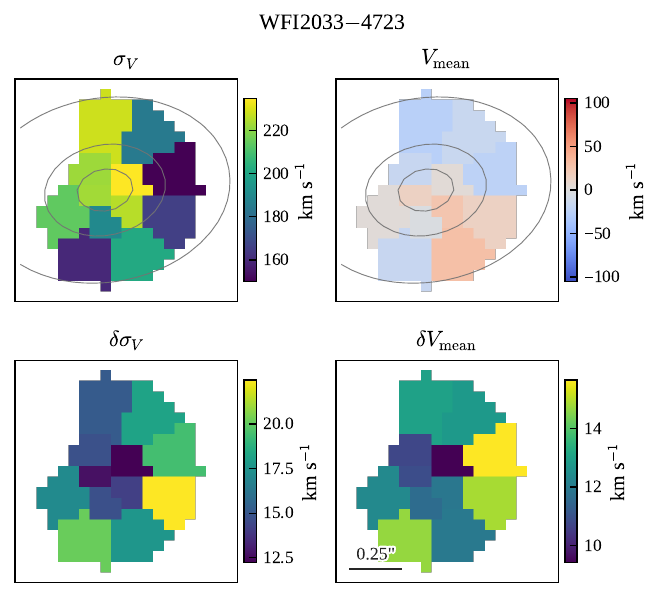}
    \caption{Same as Figure~\ref{fig:he0435_maps}.} \label{fig:wfi2033_maps}
\end{figure}

\begin{figure}[t!]
    \includegraphics[width=0.9\columnwidth]{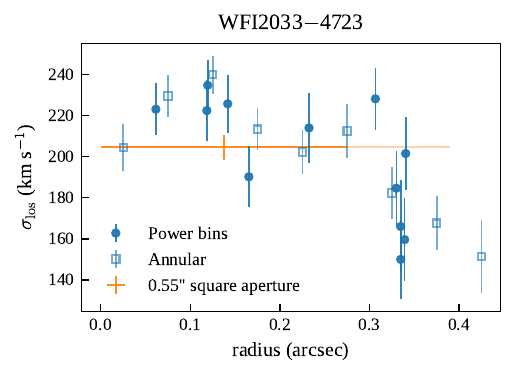}
    \caption{Same as Figure~\ref{fig:he0435_radial}.} \label{fig:wfi2033_radial}
\end{figure}

\begin{figure}[t!]
    \includegraphics[width=\columnwidth]{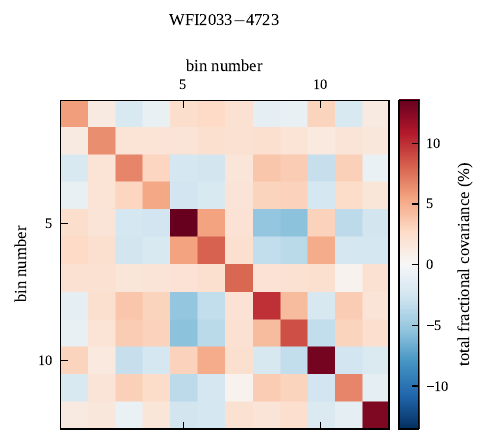}
    \caption{Same as Figure~\ref{fig:he0435_cov}.} \label{fig:wfi2033_cov}
\end{figure}

\subsection{Kinematic classification}

\begin{figure}[t!]
    \includegraphics[width=\columnwidth]{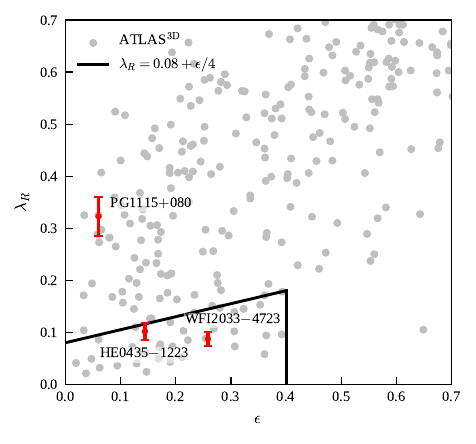}
    \caption{$\lambda_\mathrm{R}$--$\epsilon$ diagram, distinguishing fast/slow rotator classifications for ETGs using the projected specific angular momentum $\lambda_\mathrm{R}$ and the projected ellipticity $\epsilon$. The black line shows the dividing relation $\lambda_\mathrm{R} = 0.08 + \epsilon/4$ \citep{cappellari16_review}, with slow rotators below the line. Gray points show the ATLAS$^{\rm 3D}$ sample \citep{Emsellem11}.} \label{fig:lambda_r_eps}
\end{figure}

We examine the binned $V$ and $\sigma_\mathrm{v}$ maps to quantify the foreground lens deflector galaxy as a fast or slow rotator by its ellipticity $\epsilon$ and projected specific angular momentum $\lambda_\mathrm{R}$ \citep[Eqs. 5 and 6;][]{Emsellem07}, which is integrated over the kinematic maps and luminosity-weighted
by the observed \cat\ continuum flux using the same signal image used for calculating the s$S/N$ for binning (see Section~\ref{sec:binning}). These values are typically integrated within one effective radius, which we do for \lenspg\ only. We integrate within the extent of the binned spaxels for \lenshe\ and \lenswfi\ because the maximum radius is smaller than one effective radius, so that $\lambda_\mathrm{R}$ is not precisely comparable to the values given by \citet{Emsellem07} without an aperture correction. We do not apply a correction here because these classifications are primarily for understanding the expected behavior and priors for axisymmetric dynamical modeling, where high $\lambda_\mathrm{R}$ indicates strong rotational support. We list the values in Table~\ref{tab:aperture} and show them on the $\lambda_\mathrm{R}-\epsilon$ diagram in Figure~\ref{fig:lambda_r_eps}.

\lenshe\ and \lenswfi\ are slow rotators, with the caveat that $\lambda_\mathrm{R}$ is calculated within an aperture smaller than the effective radius by factors of 0.28 and 0.23, respectively. \lenshe\ appears to have a rotational axis, and it lies near the boundary of the region outlined by $\lambda_\mathrm{R} = 0.08 + \epsilon/4$, which separates the fast and slow rotator classifications, with the slow rotators below and to the left of the boundary \citep[eq.~19]{cappellari16_review}. \lenspg\ shows a clear rotation axis and outer bin $V>100$ \kmps, and its position on the $\lambda_\mathrm{R}-\epsilon$ diagram classifies it as a fast rotator.

\begin{table*}[t!]
    \centering
    \begin{tabular}{l | c c c}
        \hline
        & \lenshe\ & \lenspg\ & \lenswfi\ \\
        \hline\hline
        \multicolumn{4}{l}{\textit{Deflector}} \\
        $V_\mathrm{d}$ [\kmps]                 & \vsysdhe\ & \vsysdpg\ & \vsysdwfi\ \\
        $z_\mathrm{d}$ (published)             & \zfiddhe\ & \zfiddpg\ & \zfiddwfi\ \\
        $\Delta z_\mathrm{d}$                  & $\dzdhe$ & $\dzdpg$ & $\dzdwfi$ \\
        $z_\mathrm{d}$ (this work)             & \zdnewhe\ & \zdnewpg\ & \zdnewwfi\ \\
        \hline
        \multicolumn{4}{l}{\textit{Background source}} \\
        $V_\mathrm{host,\star}$ [\kmps]        & \vhosthe\ & \vhostpg\ & \vhostwfi\ \\
        $V_\mathrm{em}$ [\kmps]                & \vemhe\ & \vempg\ & \vemwfi\ \\
        $z_\mathrm{s}$ (published)             & \zfidshe\ & \zfidspg\ & \zfidswfi\ \\
        $\Delta z_\mathrm{s}$                  & $\dzshe$ & $\dzspg$ & $\dzswfi$ \\
        $z_\mathrm{s}$ (this work)             & \zsnewhe\ & \zsnewpg\ & \zsnewwfi\ \\
        \hline
    \end{tabular}
    \caption{Velocity offsets $V$ of the deflector and background-source
    components relative to the fiducial published redshifts, and the
    corresponding redshift adjustments (Section~\ref{sec:result}). The source
    adjustment is taken from the narrow emission-line component $V_\mathrm{em}$;
    the host stellar offset $V_\mathrm{host,\star}$ is listed for comparison
    only. For \lenshe, He~I $\lambda4026$ aligns closely with the fiducial
    redshift and is kinematically distinct from the other lines.}
    \label{tab:velocity_offsets}
\end{table*}

\subsection{Comparison with previous measurement} \label{sec:comparison_with_prev}

First products from these data were presented as single, aperture-integrated velocity dispersions in TDC-25, along with two other objects with \jwst-NIRSpec IFS. Those extracted spectra were taken from earlier data cubes reduced with a previous version of the JWST data reduction pipeline and without the benefit of the improvements achieved by the ATM models. The new reductions presented in this paper are vastly superior and have been tested for reduction systematics to more rigorous detail. For each object we extract a single integrated spectrum within a square aperture of side 0\farcs55, matching the apertures of the TDC-25 measurements, with quasar-dominated spaxels excluded and spaxel covariance propagated through the coadd. We fit this aperture spectrum identically to the spatially binned data and marginalize over the same systematics grid (see Section~\ref{sec:systematic}). We show the new values in Figures~\ref{fig:he0435_radial},~\ref{fig:pg1115_radial},~and~\ref{fig:wfi2033_radial} and list them alongside the values from TDC-25 in Table~\ref{tab:aperture}. All three are statistically consistent with the previous measurements, with differences of \dsighe, \dsigpg, and \dsigwfi\ \kmps\ for \lenshe, \lenspg, and \lenswfi, respectively. The offsets do not share a coherent direction, consistent with independent per-object improvements rather than a systematic shift. The average estimated uncertainty has been reduced from \avgunctdc\ to \avguncnew. We attribute the changes to improvements in the data reduction and to the kinematic-extraction methodology introduced and validated in this work.

\begin{table*}[t!]
    \centering
    \begin{tabular}{c|ccc|cccc}
        \hline
         Object & $\rm \sigma^{TDC-25}_{v}$ [\kmps] & $\rm \sigma_{v}$ [\kmps] & $\rm \Delta\sigma_{v}$ [\kmps] & $\lambda_R$ & $\epsilon$ & $R_{\rm max}/R_e$ & class \\
         \hline
         \lenshe\ & \sigtdche  & \sigaphe & \dsighe  & \lamrhe & \epshe & \rmaxrehe & slow$^*$ \\
         \lenspg\ & \sigtdcpg  & \sigappg & \dsigpg & \lamrpg & \epspg & \rmaxrepg & fast\\
         \lenswfi\ & \sigtdcwfi & \sigapwfi & \dsigwfi  & \lamrwfi & \epswfi & \rmaxrewfi & slow$^*$ \\
        \hline
    \end{tabular}
    \caption{Velocity dispersions measured from spectra integrated over 0\farcs55 square apertures from the data cubes presented in this work compared with the values used by \citet{tdcosmo25_milestone}, together with the luminosity-weighted rotation parameter $\lambda_R$, observed ellipticity $\epsilon$, the maximum radius probed relative to the effective radius, and the resulting rotator classification. The TDC-25 uncertainties combine the quoted statistical and systematic terms in quadrature.\\
    $^*$ --- aperture-limited $\lambda_R(<R_{\rm max})$
    } \label{tab:aperture}
\end{table*}

\section{Conclusion}
\label{sec:conclusion}

We extracted 2D stellar kinematic maps of the deflector galaxies in the quadruply imaged quasar systems \lenses, observed with \jwst-NIRSpec IFS as part of GTO program 1198 (PI: Stiavelli). The main results can be summarized as follows:
\begin{itemize}
    \item Owing to many improvements in the data reduction pipeline, calibrations, algorithms, and kinematic-extraction methodology developed by our team, the uncertainties have been reduced significantly with respect to our previous work: the average uncertainty on the aperture-integrated stellar velocity dispersions has decreased from \avgunctdc\ in TDC-25 to \avguncnew, with all three measurements statistically consistent with the previous values (within \dsigmax) and no coherent direction to the offsets.
    \item The velocity dispersion maps reach average per-bin statistical and systematic uncertainties of \statbudrange\ and \sysbudrange, respectively, with average bin-to-bin correlated errors of $\zeta_B$ \constroffdiag, meeting the accuracy requirements of precision time-delay cosmography.
    \item \lenspg\ is a fast rotator with a clear rotation axis, while \lenshe\ and \lenswfi\ are slow rotators within the probed radii, informing the priors for axisymmetric dynamical modeling.
    \item The velocity offsets of the background hosts' narrow emission lines refine the source redshifts; most notably $z_{\rm s} = \zsnewpg$ for \lenspg\ ($\Delta z_{\rm s} = \dzspg$), while the corrections for \lenshe\ and \lenswfi\ and for all deflector redshifts are small.
\end{itemize}
These data products will be used in the 2026 TDCOSMO milestone analysis to constrain cosmological parameters in combination with new lens models (\citealp{williams25_wfi2033}; Williams et al., \textit{in prep}), time delays, and line-of-sight convergence estimates (Johnson et al. 2026, submitted).
The kinematic maps and associated software will be made publicly available upon publication of this manuscript.

\begin{acknowledgments}

This work is based on observations made with the NASA/ESA/CSA James Webb Space Telescope. The data were obtained from the Mikulski Archive for Space Telescopes at the Space Telescope Science Institute, which is operated by the Association of Universities for Research in Astronomy, Inc., under NASA contract NAS 5-03127 for JWST. These observations are associated with program \#1198. The specific observations analyzed can be accessed via \url{https://dx.doi.org/10.17909/vymn-aa94}. 
TT acknowledges support by NSF through grant NSF-AST-2407277, and from the Moore Foundation through grant 8548. MS acknowledges support by  NASA grant 80NSSC22K1294. TM acknowledges the support by JSPS KAKENHI Grant Number 25K24918. Code scripts and kinematic maps are available from the corresponding author on request and will be made publicly available upon publication of this manuscript.

This research made use of \textsc{RegalJumper} and \textsc{squirrel}
\citep{shajib25b_rxj1131_nirspec}, the \textsc{jwst} calibration pipeline
\citep{bushouse26_jwst_drp_v2p0p0}, \textsc{NSClean} \citep{Rauscher24},
\textsc{lacosmic} \citep{vanDokkum01}, \textsc{raccoon} \citep{Shajib25b},
the JWST Background Tool\footnote{\url{https://github.com/spacetelescope/jwst_backgrounds}},
\textsc{pPXF} \citep{Cappellari17, cappellari23_ppxf}, \textsc{pafit}
\citep{krajnovic_06_fitkinpa}, \textsc{PowerBin} \citep{cappellari25_powerbin},
\textsc{stpsf} \citep{Perrin14}, \textsc{numpy} \citep{harris20_numpy},
\textsc{scipy} \citep{virtanen20_scipy}, \textsc{astropy} \citep{AstropyCollaboration13, AstropyCollaboration18, AstropyCollaboration22}, \textsc{pandas}
\citep{mckinney10_pandas}, \textsc{matplotlib} \citep{Hunter07},
\textsc{jupyter} \citep{Kluyver16}, \textsc{emcee} \citep{foreman13_emcee},
and \textsc{dill}\footnote{\url{https://github.com/uqfoundation/dill}}.
\end{acknowledgments}

\bibliographystyle{apsrev4-2}
\bibliography{shawn_knabel_bibliography} 

\appendix

 
\section{Pixel replacement: variants and a failure mode} \label{app:pixel_replacement}

We tested two maximally conservative approaches to pixel replacement: 1) \textit{no} replacement, and 2) a scheme where we dropped any pixel whose proposed replacement flux did not reach consensus across the reduction variants (ATM+mingrad, mingrad, and profile). Consensus is broken in the following ways: 1) the pixel was rejected (not recovered and left NaN) by any of the variants; 2) recovered flux for mingrad and profile (without ATM) disagreed to $>5\%$ on the mean of the two values; and 3) a ``flower'' shape where the orthogonal neighbors were recovered, but the central initially flagged pixel was not recovered. In practice, we ran the Spec3 pipeline through pixel replacement only, interrupting before the cube was built. We tracked NaN-transitions, and any violation of any of the consensus criteria rejected the pixel in the final consensus mask, which was applied to each of the exposures in a new \texttt{cal} file for building the consensus cubes.
 
This uncovered a failure mode that can significantly bias the extracted kinematics and, because the bias is largest where the absorption signal is strongest, preferentially affects the highest-S/N spaxels---those that would otherwise be considered the most trustworthy. Because the results of our fiducial model variant with ATM sampling are robust, this failure mode is completely avoided if one trusts that reduction, and we record it here for the sake of warning. One of the four exposures for \lenshe\ had DQ flags in the cores of two of the CaT lines in the brightest central regions of the deflector galaxy. DQ flags are applied in Stage~1 and in the following stages up to Stage~2, where outlier cleaning aggressively removes both positive and negative excursions, so the sharp core of an absorption line in a high-continuum spaxel can itself be flagged. Whether or not the flag is legitimately removing an artifact, when the core pixel of an absorption line profile is flagged with a DQ for an individual exposure, the pixel replacement is essential for ensuring the line depth is properly reconstructed. If the affected pixel is instead dropped, as in our consensus scheme or a reduction that includes no replacement, while the wings and surrounding continuum are kept, the core is sampled from only the remaining exposures while the wings and continuum are sampled from all of them. This sampling inhomogeneity biases the measured depth even when every flag is individually correct---deepening the line when, as here, the dropped exposure carries the brighter continuum at the core---and the spuriously sharpened core is fit as a lower velocity dispersion. Fitting this incorrect reduction against the fiducial with identical machinery, the fiducial was favored only at moderate significance: the two could not be cleanly separated in goodness of fit despite per-bin differences of up to $-25$\ \kmps in a central bin. We compared the CaT line depths and ratios produced by the incorrect reduction with those of the fiducial reduction and with stellar templates from both the Indo-US and XSL libraries, finding that the artificial line depths and ratios could not be proven to be non-stellar. After dropping the single affected exposure and repeating every test for the approach without pixel replacement and for our consensus scheme, the variant cubes returned to agreement with the fiducial. All results were well within the uncertainties, in both line depths and kinematics, identifying the mechanism and confirming that its removal restores consistency.

\section{Polynomial-degree stability criterion} \label{app:poly_selection}

\begin{figure*}[t!]
    \centering
    \includegraphics[width=\textwidth]{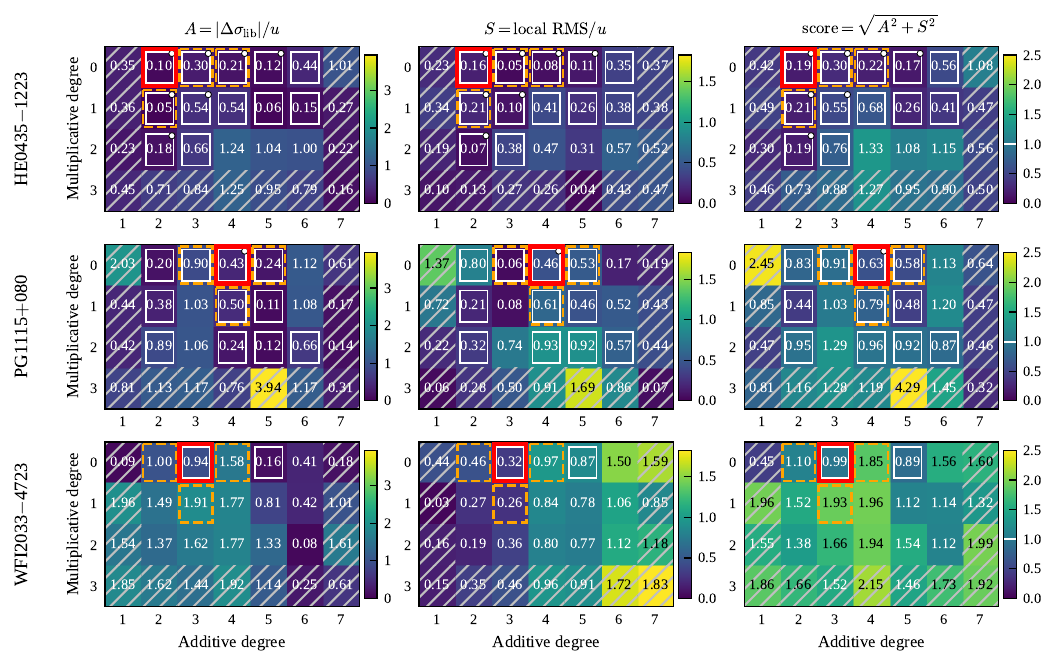}
    \caption{Polynomial-degree stability maps for the deflector fits (Section~\ref{sec:poly_selection}), one row per object. Columns: inter-library agreement $A$, neighborhood variation $S$, and the combined score $\sqrt{A^2+S^2}$, all in units of the fit uncertainty $u$; cells are the additive and multiplicative degree of the swept grid. Thin white outlines mark cells passing the score cut (score $\leq 1$, the white tick on the score color bar); white dots mark cells whose four adjacent cells also pass; the thick red outline is the fiducial and the dashed orange outlines are the candidate set carried into the systematics grid (Table~\ref{tab:systematic_choices}). Hatched cells are on the outer edge of the swept range and enter only as neighbors. \lenshe\ and \lenspg\ have contiguous passing regions; for \lenswfi\ no cell passes together with its neighborhood, and the fiducial is the lowest-degree cell that passes on its own.}
    \label{fig:poly_stability}
\end{figure*}

For each candidate cell $(n_{\rm add}, n_{\rm mult})$ in a grid of candidate degrees, we fit the aperture spectrum with both template libraries and quantify (i) the agreement between the two template libraries and (ii) the variation of the fitted dispersion across the neighboring cells, combining them into a single stability score in units of the fit uncertainty. With $u$ the rms of the two libraries' fit uncertainties at that cell, the agreement term is $A = |\sigma_{\rm IU}-\sigma_{\rm XSL}|/u$ and the neighborhood term $S$ is the rms deviation of the two-library mean dispersion from its values in the adjacent cells $(n_{\rm add}\pm1, n_{\rm mult})$ and $(n_{\rm add}, n_{\rm mult}\pm1)$ that lie within the swept grid, divided by $u$; the score is $\sqrt{A^2+S^2}$, and a cell passes if its score is $\leq 1$. For the deflector fits the grid spans additive degrees \gridlensadd\ and multiplicative degrees \gridlensmult\ for all three objects, with the lowest and highest additive degrees and the highest multiplicative degree serving only as neighbors; for the host fits the grid is chosen per object to bracket the stable region (\gridhosthe, \gridhostpg, and \gridhostwfi\ for \lenshe, \lenspg, and \lenswfi, respectively). We select the lowest-degree cell whose score passes together with its neighborhood. The candidate set carried into the systematics grid (Table~\ref{tab:systematic_choices}) is the contiguous neighborhood of the fiducial. For \lenswfi, no cell passes the neighborhood criterion, and the fiducial is taken as the lowest-degree cell that passes the score cut on its own, with the neighborhood requirement relaxed; the surrounding candidates are admitted to the systematic grid without the passing cut. The stability maps are shown in Figure~\ref{fig:poly_stability}.

\section{Outlier identification} \label{app:spikes}

We identify spurious features from the residuals after an initial fit, normalized by the per-pixel noise, with an iterative rejection adapted from the generalized Extreme Studentized Deviate test \citep[ESD;][]{Rosner1983}: at each iteration the residuals are studentized by the robust (MAD) scatter of the remaining pixels and the most extreme pixel is removed if it exceeds a fixed threshold, with rejection capped at 5\% of the fitted pixels. The minimum threshold for rejection is set from a principled value where the expected number of pixels with absolute residuals larger than the threshold is less than 1, given the number of data points in the fit window ($\sim2.5\sigma$, see Table~\ref{tab:systematic_choices}). We first fit all wavelength pixels in the spectrum with the given combination of ingredients and perform this rejection on the residual. We perform this procedure with both template libraries and reject any wavelength pixels identified in \textit{either} of the two fits; identical masks are then applied to the final fits conducted with both libraries. This ensures the libraries are compared consistently, since different masks would result in fundamentally different datasets for comparison. To prevent the rejection step from clipping genuine kinematic signal, the cores of the three \cat\ lines are protected: the predicted core positions are computed from each bin's fitted velocity, and the nearest detector pixel $\pm1$ pixel around each core is exempt from rejection.

TDC-XXIV considered some of these features to be from astrophysical sources at different redshifts without a detectable continuum and modeled them explicitly as additional narrow emission-line components at instrumental resolution. We tested that approach here and do not adopt it for our data. The additional free components can absorb the shape of the \cat\ lines in addition to fitting isolated artifacts, and for our lowest-S/N object, \lenswfi, they biased the recovered aperture velocity dispersion upward by $\sim15\%$ while improving the goodness of fit, so that BIC-weighting over the two outlier methods would have favored the biased fits. For \lenshe\ and \lenspg\ the two treatments agree within the statistical uncertainties. Masking removes the same pixels without that freedom, and we use it exclusively.


    
\begin{figure*}[t!]
	\includegraphics[width=\textwidth,height=0.82\textheight,keepaspectratio]{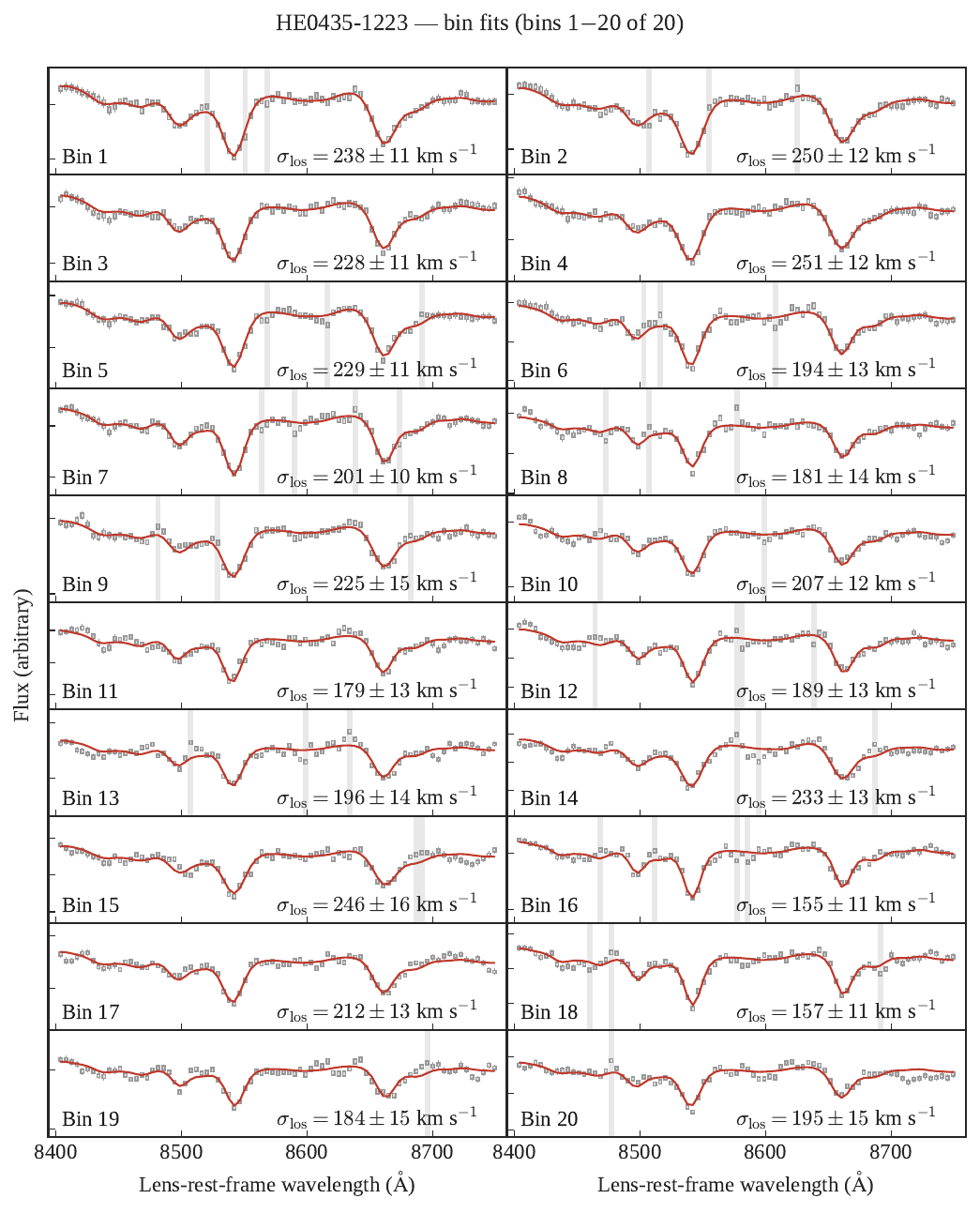}
	\caption{Kinematic fits to the individual Power bin spectra of \lenshe. Gray rectangles show the data and $1\sigma$ uncertainty. Small vertical lines behind the data show the inflated uncertainty for pixels near the edges of the fitted spectra. Red curves are the best-fit model, and gray vertical bands are the pixels masked by the outlier rejection (Section~\ref{sec:spikes}). The measured velocity dispersion is annotated in each panel.
    \label{fig:he0435_bin_fits}
	}
\end{figure*}

\section{Fits of individual bins} \label{app:individual_voronoi_bin_fits}

\begin{figure*}[t!]
	\includegraphics[width=\textwidth,height=0.9\textheight,keepaspectratio]{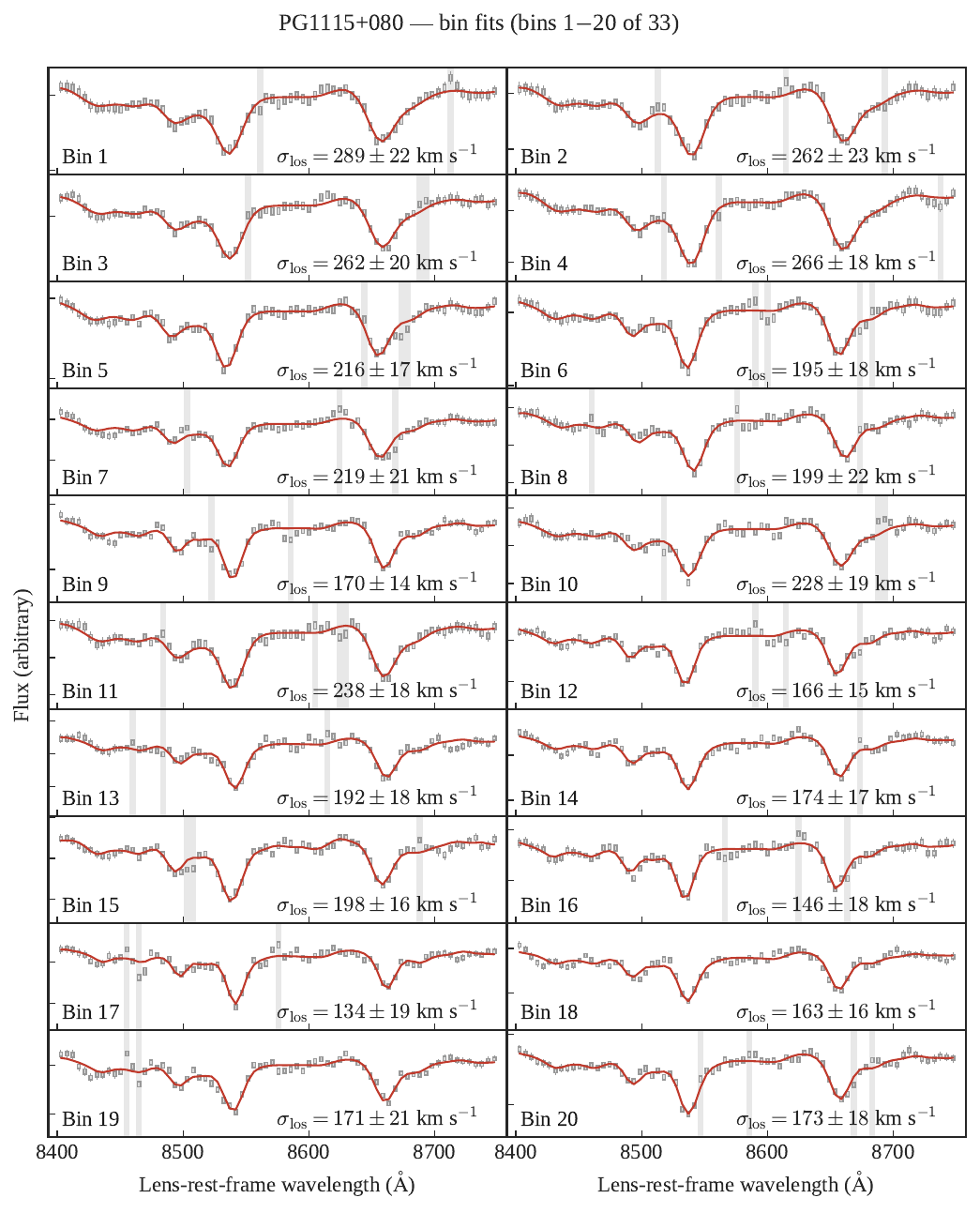}
	\caption{Same as Figure~\ref{fig:he0435_bin_fits} for \lenspg\ (bins 1--20).\label{fig:pg1115_bin_fits_1}}
\end{figure*}

\begin{figure*}[t!]
	\includegraphics[width=\textwidth,height=0.9\textheight,keepaspectratio]{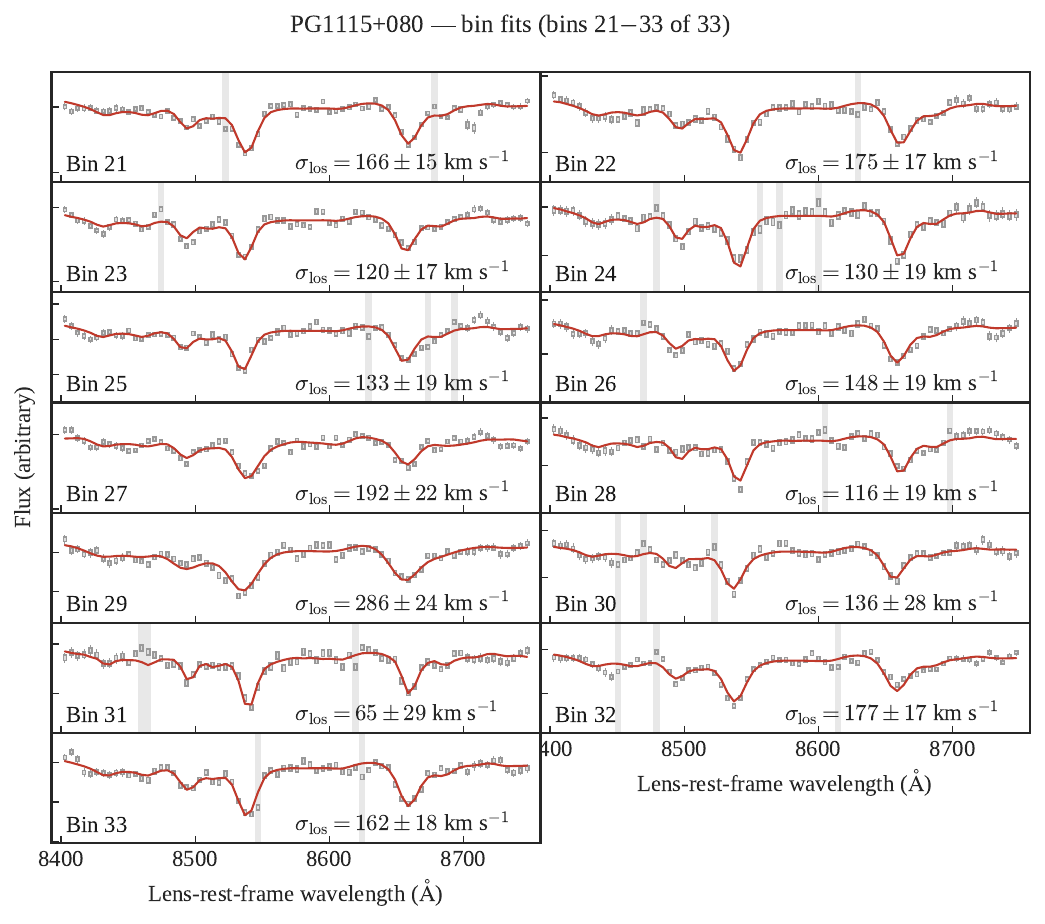}
	\caption{Same as Figure~\ref{fig:he0435_bin_fits} for \lenspg\ (bins 21--33).\label{fig:pg1115_bin_fits_2}
		}
\end{figure*}

\begin{figure*}[t!]
	\includegraphics[width=\textwidth,height=0.9\textheight,keepaspectratio]{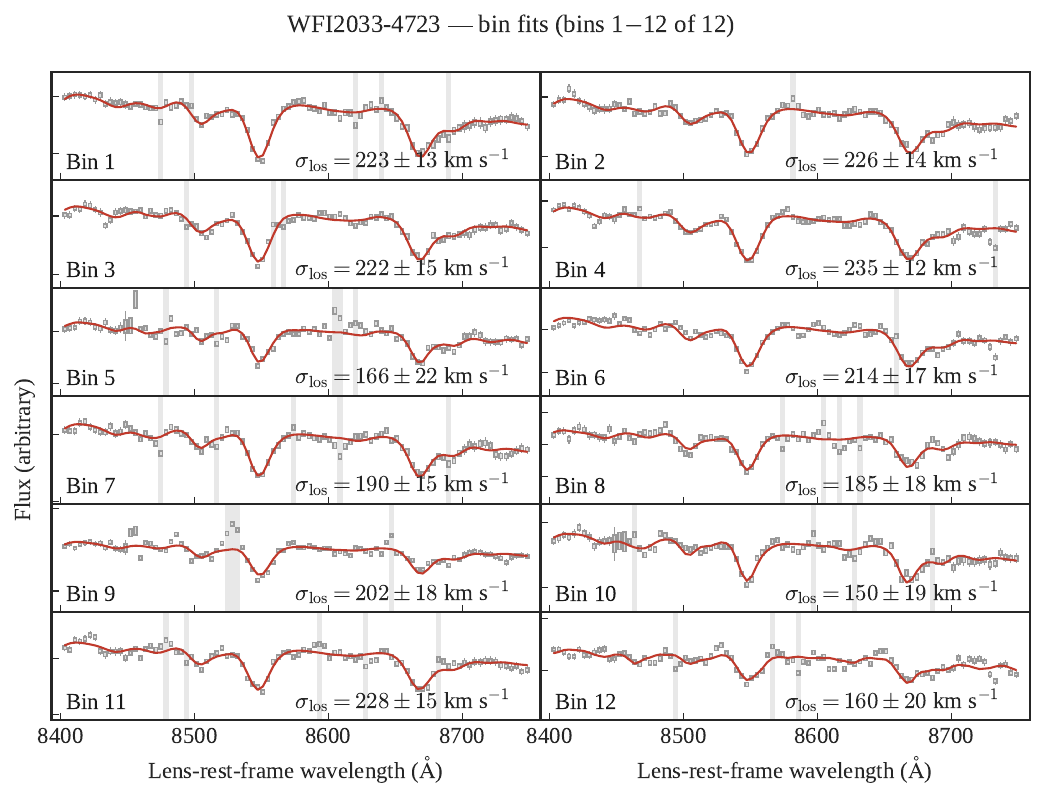}
	\caption{Same as Figure~\ref{fig:he0435_bin_fits} for \lenswfi\ (bins 1--12).\label{fig:wfi2033_bin_fits_1}}
\end{figure*}

In this appendix, we illustrate the kinematic fits to all the individual spatially-binned spectra in Figures~\ref{fig:he0435_bin_fits}--\ref{fig:wfi2033_bin_fits_1}.


\end{document}